\documentclass[%
 reprint,
superscriptaddress,
 amsmath,amssymb,
 aps,
 pra,
]{revtex4-2}

\usepackage{dcolumn}
\usepackage[colorlinks=true, allcolors=blue]{hyperref}
\usepackage{braket}
\usepackage{bm}
\usepackage[capitalize]{cleveref}
\usepackage{amsfonts}
\usepackage{mathrsfs}
\usepackage{mathtools}
\usepackage{amsthm}
\usepackage{graphicx}
\usepackage{color}
\usepackage{array}
\usepackage[makeroom]{cancel}
\usepackage{changes}
\usepackage{bbm}
\usepackage{comment}
\usepackage{multirow}
\usepackage{cellspace} 
\usepackage{appendix}
\usepackage{mdframed}
\usepackage{algorithm}
\usepackage{algpseudocode}
\usetikzlibrary{positioning, shapes, arrows, fit}
\usetikzlibrary{arrows.meta}
\usepackage{orcidlink}
\usepackage{leftindex}
\usepackage{empheq}
\usepackage{booktabs}
\usepackage{afterpage}
\usepackage{xcolor}
\usepackage{tikz}
\usetikzlibrary{quantikz2}

\crefname{thm}{Theorem}{Theorems}
\crefname{dfn}{Definition}{Definitions}
\crefname{rmk}{Remark}{Remarks}
\crefname{lem}{Lemma}{Lemmas}
\crefname{cor}{Corollary}{Corollaries}

\theoremstyle{plain}

\theoremstyle{remark}

\newcommand{\rixs}{\ket{R_{\bm{\varepsilon}_I, \bm{\varepsilon}_S}(\omega_I)}}
\newcommand{\floor}[1]{\left\lfloor #1 \right\rfloor}
\newcommand{\ceil}[1]{\left\lceil #1 \right\rceil}

\usepackage[english]{babel}
\usepackage[utf8]{inputenc}

\definecolor{X1}{HTML}{4D53C8}
\definecolor{X2}{HTML}{52BD7C}
\definecolor{X3}{HTML}{44ACE8}
\definecolor{X4}{HTML}{D7333B}

\begin{document}
\title{Quantum algorithm for simulating resonant inelastic X-ray scattering in battery materials}

\author{Ignacio Loaiza \orcidlink{0000-0001-9630-855X}}
\affiliation{Xanadu, Toronto, ON, M5G 2C8, Canada}

\author{Alexander Kunitsa \orcidlink{0000-0002-3640-8548}}
\affiliation{Xanadu, Toronto, ON, M5G 2C8, Canada}

\author{Stepan Fomichev \orcidlink{0000-0002-1622-9382}}
\affiliation{Xanadu, Toronto, ON, M5G 2C8, Canada}

\author{Danial Motlagh \orcidlink{0009-0003-7655-4341}}
\affiliation{Xanadu, Toronto, ON, M5G 2C8, Canada}

\author{Diksha  Dhawan \orcidlink{0000-0002-1129-3166}}
\affiliation{Xanadu, Toronto, ON, M5G 2C8, Canada}

\author{Soran Jahangiri \orcidlink{0000-0002-9988-8841}}
\affiliation{Xanadu, Toronto, ON, M5G 2C8, Canada}

\author{Juliane Holst Fuglsbjerg \orcidlink{}}
\affiliation{Department of Chemistry, University of Copenhagen, DK-2100 Copenhagen Ø, Denmark}

\author{Artur F. Izmaylov \orcidlink{0000-0001-8035-6020}}
\affiliation{Department of Physical and Environmental Sciences, University of Toronto Scarborough, Toronto, Ontario M1C 1A4, Canada}

\author{Nathan Wiebe \orcidlink{0000-0001-6047-0547}}
\affiliation{University of Toronto, Toronto, Canada}

\author{Yaser Abu-Lebdeh \orcidlink{0000-0001-8936-4238}}
\affiliation{Clean Energy Innovation Research Centre, National Research Council of Canada, Ottawa K1A 0R6, Canada}

\author{Juan Miguel Arrazola \orcidlink{0000-0002-0619-9650}}   
\affiliation{Xanadu, Toronto, ON, M5G 2C8, Canada}

\author{Alain Delgado \orcidlink{0000-0003-4225-6904}}
\affiliation{Xanadu, Toronto, ON, M5G 2C8, Canada}

\begin{abstract}
Resonant inelastic X-ray scattering (RIXS) is the workhorse experimental technique for probing the structural degradation of higher-capacity cathode materials. However, the interpretation of experimental spectra is challenging due to the lack of accurate simulations. In this work, we propose a quantum algorithm for simulating the RIXS spectrum of molecular clusters hypothesized to form in Li-excess cathodes. The algorithm uses quantum phase estimation to sample the spectrum from a state encoding the scattering transition amplitudes of the cluster valence excitations. We prepare this state in the quantum computer using a block-encoding of the dipole operator and quantum signal processing to implement the Green's function propagator over intermediate core-excited states. To showcase the algorithm, we use a model cluster proposed in recent experimental works consisting of an oxygen dimer bonded to a manganese atom. Using the PennyLane software platform, we report resource estimation for simulating RIXS spectra for chemically motivated active spaces of increasing sizes. For a classically challenging active space with 20 orbitals, the algorithm requires $2.0 \times 10^{10}$ Toffoli gates and $414$ logical qubits.
\end{abstract}

\maketitle

\section{Introduction}
\label{sec:intro}
Lithium-excess cathode materials such as the archetype $\text{Li}_{1.2}\text{Ni}_{0.13}\text{Co}_{0.13}\text{Mn}_{0.54}\text{O}_2$ (Li-rich NMC) offer a promising avenue to achieve cell-level energy densities greater than 500 Wh/kg, a key target for wide adoption of electric vehicles~\cite{zhang2022pushing, liu2022battery500}. This comes from their excess capacity due to the anion redox activity~\cite{seo2016structural}. However, irreversible structural changes in the cathode over just a few charge-discharge cycles lead to voltage hysteresis and rapid capacity fade, preventing their commercialization~\cite{zhang2022pushing}.
Despite previous experimental and theoretical work suggesting that molecular oxygen formation and oxygen release are the root causes of the cathode degradation~\cite{mccoll2022transition, house2023delocalized, eum2020voltage, boivin2021role, ben2019unified}, the precise nature of the local structural transformations remains a subject of intense debate~\cite{gao2025clarifying}.
\begin{figure*}
    \centering
    \includegraphics[width=\textwidth]{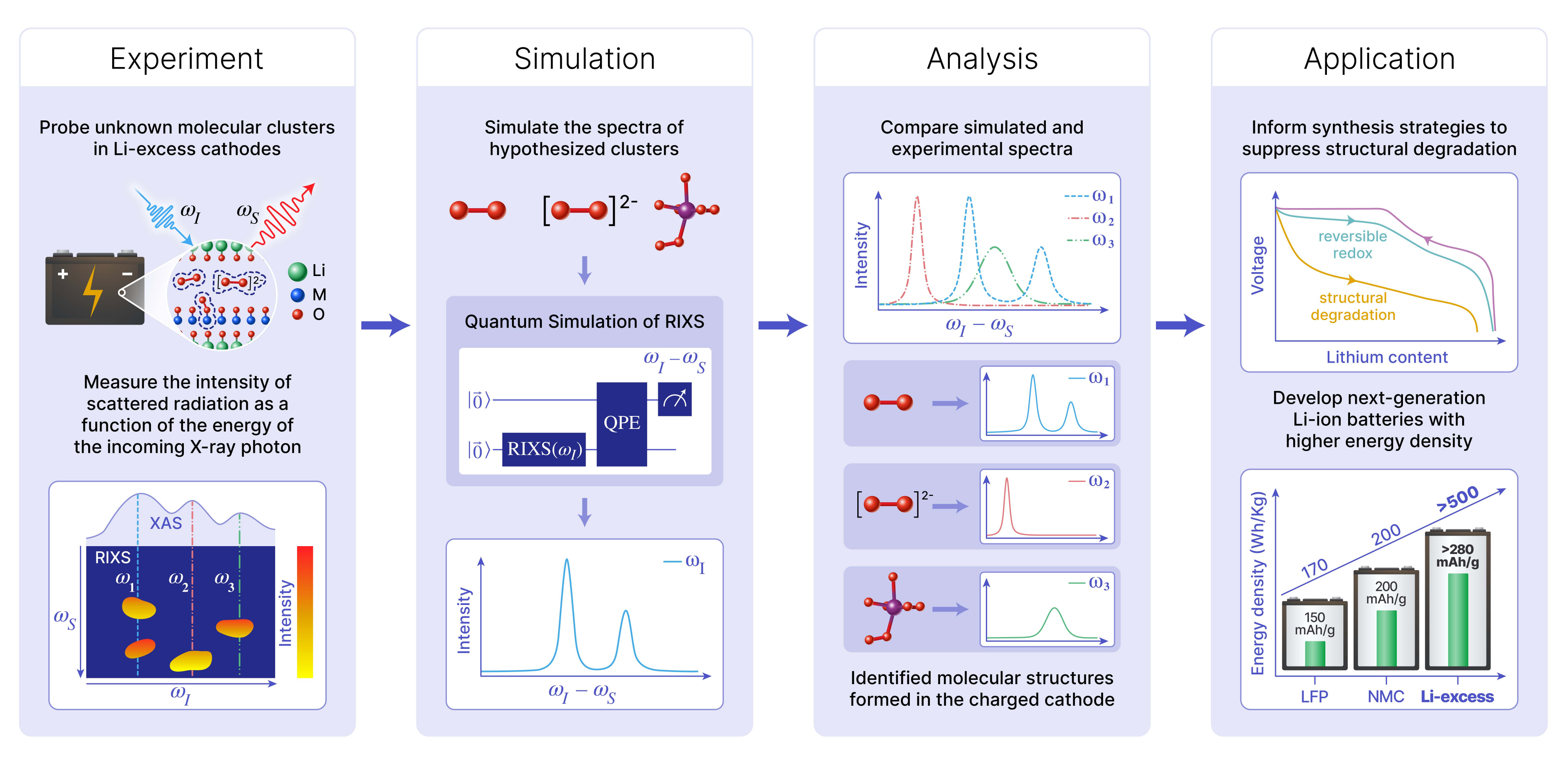}
\caption{A high-level description of the application pipeline for quantum simulation of RIXS spectra for structural analysis of lithium-excess cathode materials. First, experimental RIXS maps are built by measuring the intensity of scattered X-ray photons while scanning over the energies $\omega_I$ of the incoming radiation. For selected values of $\omega_I$ corresponding to resonances in the X-ray absorption spectrum (XAS), high-resolution RIXS spectra reveal the low-lying electronic excited states of the molecular clusters in the charged cathode. To interpret the spectra, we accurately simulate the RIXS cross-section of hypothesized structures -- such as oxygen dimers and clusters with transition metals -- using a quantum computer. The quantum algorithm prepares the initial state $\ket{\text{RIXS}(\omega_I)}$ and uses quantum phase estimation to sample the valence excited states of the cluster with probability given by the RIXS transition amplitude. The experimental spectra are compared with theoretical simulations of hypothesized clusters to fingerprint their RIXS features in the experimental spectra. This analysis provides a pathway to understand the causes of structural degradation in Li-excess cathodes. This is critical for informing new synthesis strategies to stabilize the structure of these materials to enable new battery cells with higher energy density.}
\label{fig:hero}
\end{figure*}

Resonant X-ray inelastic scattering spectroscopy (RIXS) has emerged as the preeminent technique for probing structural degradation in Li-excess cathodes~\cite{gao2025clarifying, house2023delocalized, xu2018elucidating}. The main focus of applying RIXS to Li-excess cathodes has been to investigate the excited states of the oxygen species that are hypothesized to be central to the degradation mechanism. The leading interpretation of the observed RIXS spectra is that the dominant peaks correspond to vibrational and electronic excitations of \textit{molecular oxygen}, supposedly forming during the charging cycles~\cite{house2023delocalized, mccoll2022transition, hirsbrunner2024vibrationally}. However, very recently Gao {\it et al.}~\cite{gao2025clarifying} demonstrated that the RIXS features linked to molecular oxygen formation in Li-excess also appear in conventional cathodes, and suggested instead that more complicated molecular clusters, involving oxygen dimers bonded to transition metal species in the material, may form in the delithiated cathode. This is strong evidence that more effort is required to understand the relationship between the redox processes and structural transformations in Li-excess materials. Importantly, having access to accurate first-principles simulations of the RIXS spectra of the molecular clusters hypothesized to form in these materials is key for a more reliable interpretation of RIXS experiments~\cite{zhang2022pushing}.

Accurately simulating RIXS spectra of Li-excess materials is challenging for classical methods. The systems of interest, consisting of oxygen dimers bonded to transition metals, are typically strongly correlated~\cite{gao2025clarifying, EPR2013reversible}. For this reason, accurate simulations of the RIXS cross-section must rely on multi-reference wavefunction-based methods for a balanced description of the ground, intermediate, and final excited states involved in the scattering process~\cite{rixs_review_nature}. At the O K-edge, the intermediate states correspond to electronic excitations between O $1s$ core and valence states mixing $p$ and $d$ orbitals of the oxygen dimer and the nearest-neighbour transition metals, respectively. This makes the calculations of intermediate states classically challenging despite the relatively small sizes of these clusters~\cite{fomichev2025fast}. Since RIXS is a second-order process, we also require an accurate description of the final states consisting of valence-valence excitations between frontier orbitals, which are more delocalized. This often requires using large active spaces that exceed the capabilities of complete active space methods, rendering classical RIXS simulations infeasible~\cite{lee2023ab, maganas2014combined}.

In this work, we advance recent research on quantum simulations for first-order spectroscopy of battery materials \cite{fomichev2025fast, kunitsa2025quantum} by proposing a quantum algorithm to simulate RIXS --- a second-order spectroscopy --- and applying it to molecular clusters in Li-excess materials. The algorithm is designed to be a core element within the workflow for interpreting experimental RIXS spectra of battery materials, as outlined in Fig. \ref{fig:hero}. We simulate the RIXS spectrum by encoding the Kramers-Heisenberg amplitudes of the RIXS process \cite{rixs_review_nature} for a specific frequency of the incoming photons in a RIXS initial state, which we then sample with quantum phase estimation (QPE). 

The initial RIXS state is prepared by applying a sequence of three operators to the cluster's ground state. First, the dipole operator is applied to create a superposition over core-excited states; the second is a Green's function propagator filtering out the off-resonant intermediate states, implemented using generalized quantum signal processing (GQSP) \cite{gqsp}; and the final is another application of dipole operator that fills the core hole, giving the desired scattering superposition over the final (valence) excited states. This three-step process constitutes a block-encoding of the RIXS state, which is then amplitude amplified to ensure success, mitigating the sampling complexity cost of the algorithm. We use the qubitized walk operator from the BLISS-THC method \cite{bliss_thc,thc} for both GQSP and QPE subroutines. Applying this algorithm to simulate the RIXS spectra of a transition-metal oxide cluster ($\text{MnO}_7\text{H}_6$) predicted to form in Li-excess cathodes \cite{gao2025clarifying}, we find that for a classically challenging active space with 20 spatial orbitals, the algorithm requires $414$ logical qubits and $2.0 \times 10^{10}$ Toffoli gates. These resource estimates were computed using the PennyLane software platform; a comprehensive summary of resource estimates for varying system sizes is given in \cref{tab:re_rixs}. 
\begin{table*}[t]
\centering
%
%
%
%
%
%
%
%
%
%
\begin{tabular}{c c c c c c}
\hline \addlinespace[0.05 cm]
$N_e$ &~~~~$N_a$ &~~~~\text{FCI matrix dimension} &~~~~1-norm &~~~~Logical qubits &~~~~Toffoli gates \\
\addlinespace[0.05 cm] \hline \addlinespace[0.1 cm]
$15$ &~~~~$16$ &~~~~ $1.5 \times 10^8$    &~~~~$105.37$ &~~~~ $351$ &~~~~ $1.38 \times 10^{10}$ \\
$19$ &~~~~$18$ &~~~~ $2.1 \times 10^9$    &~~~~$117.46$ &~~~~ $384$ &~~~~ $1.68 \times 10^{10}$ \\
$19$ &~~~~$20$ &~~~~ $3.1 \times 10^{10}$ &~~~~$125.51$ &~~~~ $414$ &~~~~ $2.00 \times 10^{10}$ \\
$21$ &~~~~$22$ &~~~~ $4.6 \times 10^{11}$ &~~~~$141.43$ &~~~~ $449$ &~~~~ $2.55 \times 10^{10}$ \\
$21$ &~~~~$24$ &~~~~ $4.9 \times 10^{12}$ &~~~~$148.47$ &~~~~ $479$ &~~~~ $2.93 \times 10^{10}$ \\
$21$ &~~~~$26$ &~~~~ $4.1 \times 10^{13}$ &~~~~$166.23$ &~~~~ $509$ &~~~~ $3.59 \times 10^{10}$ \\
$23$ &~~~~$28$ &~~~~ $6.5 \times 10^{14}$ &~~~~$160.45$ &~~~~ $539$ &~~~~ $3.72 \times 10^{10}$ \\
$27$ &~~~~$30$ &~~~~ $1.7 \times 10^{16}$ &~~~~$205.65$ &~~~~ $570$ &~~~~ $5.25 \times 10^{10}$ \\
\hline
\end{tabular}
\caption{Resource estimates for simulating the RIXS spectrum of the molecular cluster detailed in \Cref{sec:aplication} as we increase the number of orbitals $N_a$ and electrons $N_e$ in the active space. For the sake of completeness, the dimension of the full configuration interaction (FCI) Hamiltonian matrix and the 1-norm of the Hamiltonian are given. In all cases, we consider a reference state with total spin projection $S_z=1/2$. The reported number of logical qubits and Toffoli gates corresponds to one shot of the quantum algorithm. The sampling cost of recovering the entire spectrum is estimated as $2 \times 10^3$, as determined to recover the RIXS spectrum to the desired accuracy through QPE sampling. The code used to calculate the presented estimates uses PennyLane's resource estimator module \cite{JaySoni2026}.}\label{tab:re_rixs}
\end{table*}

The rest of the manuscript is structured as follows: in \cref{sec:rixs} we provide the theoretical framework for computing the RIXS cross-section, define the observables required for quantum simulations and analyze the limitations of state-of-the-art classical approaches. \cref{sec:algo} describes the end-to-end quantum algorithm for simulating RIXS spectra, as well as the optimizations implemented to reduce the cost of quantum simulations. In~\cref{sec:aplication} we showcase the algorithm using a prototypical cluster hypothesized to form in Li-excess materials, and perform resource estimation for quantum simulations of classically intractable active spaces. Finally, the main conclusions of the work are discussed in~\cref{sec:conclusions}.

\section{Resonant inelastic X-ray scattering}
\label{sec:rixs}
In this section we provide the theoretical framework to compute the RIXS cross-section and define the observables required for quantum simulations. In addition, we discuss the limitations of classical methods for simulating the RIXS spectrum of molecules.

\subsection{Theoretical background}
\label{ssec:theory}
In a RIXS experiment, the material is illuminated by an X-ray photon with energy $\omega_I$ near the absorption edge, wave vector $\bm{k}_I$ and polarization $\bm{\varepsilon}_I$. The scattered photon, with $\omega_S$, $\bm{k}_S$, and $\bm{\varepsilon}_S$, is detected to measure the energy loss $\omega = \omega_I - \omega_S$ to probe the materials' excitations.

\begin{figure}[b]
    \centering
    \includegraphics[width=1.0\linewidth]{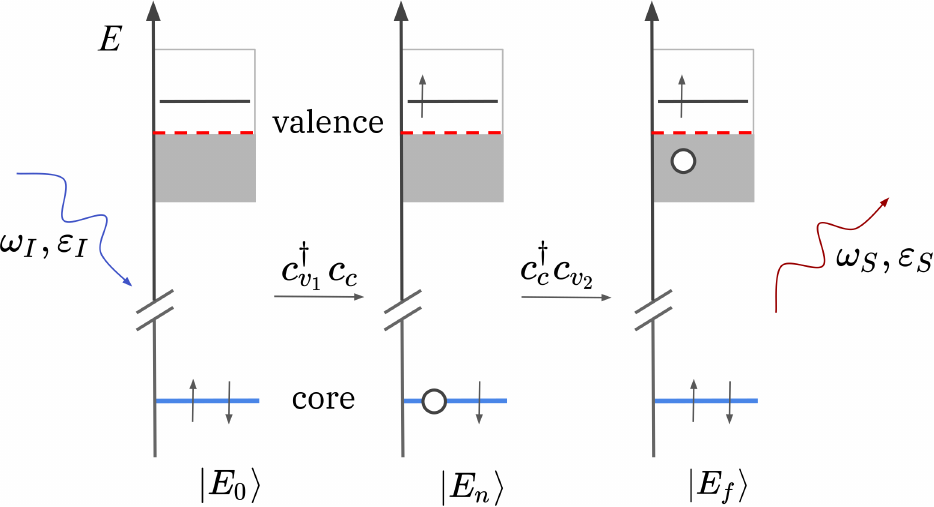}
    \caption{Sketch illustrating the two-step RIXS process in a molecular cluster. First, a photon with energy $\omega_I$ and polarization $\bm{\varepsilon}_I$ is annihilated, and a core electron is excited to a valence orbital ($\hat{c}_{v_1}^\dagger \hat{c}_c$). This interaction drives a resonant transition between the (many-body) ground $\ket{E_0}$ and intermediate states $\ket{E_n}$ of the electronic system. Then, a valence electron below the Fermi level (red dashed line) transitions to fill the core hole ($\hat{c}_c^\dagger\hat{c}_{v_2}$) leaving the cluster in a valence excited state $\ket{E_f}$, and the scattered photon with energy $\omega_s$ and polarization $\bm{\varepsilon}_S$ is created.}
    \label{fig:rixs_states}
\end{figure}

RIXS is a two-step process described within second-order perturbation theory applied to the light-matter interaction Hamiltonian~\cite{rixs-rmp-2011}. As illustrated in~\cref{fig:rixs_states}, an incoming photon is annihilated and a core electron is promoted to an empty band (orbital) of the material (molecule).
This takes the electronic system from the ground $\ket{E_0}$ to an intermediate state $\ket{E_n}$ carrying a core hole. Then, the core hole is filled by a valence electron, leaving the system in a final state $\ket{E_f}$ corresponding to a valence excitation, accompanied by the creation of the outgoing (scattered) photon. The transition amplitude of this process is given by~\cite{rixs_review_nature}
\begin{align}
& W_{f0}[(\omega_I, \bm{k}_I,\bm{\varepsilon}_I), (\omega_S, \bm{k}_S, \bm{\varepsilon}_S)] \nonumber \\
& = \sum_n \frac{ \bra{E_f} \hat{\mathcal{D}}^\dagger(\omega_S, \bm{k}_S, \bm{\varepsilon}_S) \ket{E_n} \bra{E_n} \hat{\mathcal{D}}(\omega_I, \bm{k}_I, \bm{\varepsilon}_I) \ket{E_0}}{\omega_I - (E_n-E_0)+i\Gamma},
\label{eq:rixs_amplitude}
\end{align}
where $E_0$ and $E_n$ are, respectively, the energies of the ground and intermediate states, and $\Gamma$ is the inverse of the intermediate state lifetime in units of energy. The light-matter interaction operator $\hat{\mathcal{D}}$ is defined as~\cite{rixs-rmp-2011}:
\begin{equation}
\hat{\mathcal{D}}(\omega, \bm{k}, \bm{\varepsilon}) = \frac{1}{im\omega} \sum_{j=1}^{N_e} e^{i\bm{k}\cdot\bm{r}_j} \bm{\varepsilon} \cdot \hat{\bm{p}}_j,
\label{eq:lm_int}
\end{equation}
where $m$ is the electron mass, $N_e$ is the number of electrons and $\bm{p}$ is the momentum operator. For wavelengths larger than the radial extent of the core orbital, one can assume the dipole limit $e^{i\bm{k}\cdot\bm{r}_j} \approx 1$~\cite{rixs-rmp-2011}. This removes the momentum dependence, simplifying the operator $\mathcal{D}$ to the equation
\begin{equation}
\hat{D}_\varepsilon = \sum_{j=1}^{N_e} \bm{\varepsilon} \cdot \bm{r}_j,
\label{eq:D_simp}
\end{equation}
and allowing us to write \cref{eq:rixs_amplitude} as
\begin{align}
W_{f0}(\omega_I, \bm{\varepsilon}_I, \bm{\varepsilon}_S)
= \sum_n \frac{ \bra{E_f} \hat{D}_{\bm{\varepsilon}_S}^\dagger \ket{E_n} \bra{E_n} \hat{D}_{\bm{\varepsilon}_I} \ket{E_0}}{\omega_I - (E_n-E_0)+i\Gamma}.
\label{eq:rixs_amplitude_simp}
\end{align}
The double-differential cross-section for the photon energy loss $\omega=\omega_I-\omega_S$ is given by~\cite{rixs-rmp-2011}
\begin{align}
&\frac{\partial^2\sigma}{\partial\Omega \partial\omega} \propto \nonumber \\
& \kern+10pt \omega_S^3 \omega_I \underbrace{\sum_f \left| W_{f0}(\omega_I, \bm{\varepsilon}_I, \bm{\varepsilon}_S) \right|^2 \delta(\omega-(E_f-E_0))}_{\equiv P_{\bm\varepsilon_I,\bm\varepsilon_S}(\omega_I,\omega)},
\label{eq:rixs_cross_section}
\end{align}
where $E_f$ is the energy of the final states, and the Dirac delta function enforces energy conservation. Here we have defined the RIXS amplitude $P_{\bm\varepsilon_I,\bm\varepsilon_S}(\omega_I,\omega)$ which completely determines the double-differential cross-section, and can be written as 
\begin{align}
\kern-8pt P_{\bm\varepsilon_I,\bm\varepsilon_S}&(\omega_I,\omega) = \nonumber\\
&\sum_f \vert\bra{E_f} \hat R_{\bm\varepsilon_I,\bm\varepsilon_S}(\omega_I)\ket{E_0}\vert^2 \delta(\omega-(E_f-E_0)).
\label{eq:rixs_prob_amplitude}
\end{align}
The operator $\hat R_{\bm\varepsilon_I,\bm\varepsilon_S}(\omega_I)$ is given by~\cite{rixs_review_nature}
\begin{equation}
\hat{R}_{\bm{\varepsilon}_I, \bm{\varepsilon}_S}(\omega_I) = \hat{D}_{\bm{\varepsilon}_S}^\dagger \hat{G}(\omega_I, \Gamma) \hat{D}_{\bm{\varepsilon}_I},
\label{eq:rixs_operator}
\end{equation}
where $\hat{G}(\omega, \Gamma)$ is the Green's function
\begin{equation}
\hat{G}(\omega, \Gamma) = \frac{1}{\omega-(\hat{H}-E_0) + i\Gamma}.
\label{eq:green_function}
\end{equation}
In~\cref{eq:green_function}, $\hat{H}$ is the electronic Hamiltonian of the system describing both the valence and core electrons. Note that the operator $\hat R_{\bm{\varepsilon}_I, \bm{\varepsilon}_S} (\omega_I)$ acting on the ground state describes the absorption of a photon with energy $\omega_I$, the propagation of the photo-excited system, and the subsequent emission of a photon with energy $\omega_S$. This is a key component of the quantum algorithm described in the next section.

The Hamiltonian $\hat{H}$ is represented in second quantization in a basis consisting of the $N_a$ spatial orbitals included in the active space
\begin{align}
\hat{H} = E^0 &+ \sum_{p, q=1}^{N_a} \sum_{\sigma \in\{\uparrow, \downarrow\}} h_{pq} \hat{c}_{p\sigma}^\dagger \hat{c}_{q\sigma} \nonumber \\
&+ \frac{1}{2} \sum_{p, q, r, s=1}^{N_a}\sum_{\sigma, \sigma'\in\{\uparrow, \downarrow\}} V_{pqrs} \hat{c}_{p\sigma}^\dagger \hat{c}_{q\sigma} \hat{c}_{r\sigma'}^\dagger \hat{c}_{s\sigma'},
\label{eq:hamiltonian}
\end{align}
where $\hat{c}$ and $\hat{c}^\dagger$ are the electron annihilation and creation operators, respectively, and $E^0$ is the energy contribution due to the frozen doubly-occupied (docc) orbitals
\begin{equation}
E^0 = \sum_{i=1}^{N_\text{docc}} \left( 2h_{ii} + \sum_{j=1}^{N_\text{docc}} 2V_{iijj} - V_{ijji}  \right).
\label{eq:fzc_energy}
\end{equation}
The coefficients $V_{pqrs}$ are the two-electron integral defined in chemists' notation as
\begin{equation}
V_{pqrs} = \int d\bm{r}_1 d\bm{r}_2~\phi_p^*(\bm{r}_1) \phi_q(\bm{r}_1) \frac{1}{|\bm{r}_1-\bm{r}_2|} \phi_r^*(\bm{r}_2) \phi_s(\bm{r}_2),
\label{eq:2e_integrals}
\end{equation}
and
\begin{align}
h_{pq} =& \int d\bm{r}~\phi_p^*(\bm{r}) \left( -\frac{\nabla^2}{2} + U(\bm{r}) \right) \phi_q(\bm{r}) \nonumber \\
&+ \sum_{i=1}^{N_\text{docc}} 2V_{pqii}-V_{piiq},
\label{eq:1e_integrals}
\end{align}
where $U(\bm{r})$ is the nuclei Coulomb potential, and $\phi_i(\bm{r})$ is the $i$th spatial orbital. Similarly, the second-quantized representation of the dipole operator in~\cref{eq:D_simp} is given by
\begin{equation}
\hat D_{\bm\varepsilon} = \sum_{p,q=1}^{N_a} \sum_{\sigma \in\{\uparrow, \downarrow\}} d^{(\bm\varepsilon)}_{pq} \hat c^\dagger_{p\sigma} \hat c_{q\sigma},
\label{eq:dip_sq}
\end{equation}
where
\begin{equation}
d_{pq}^{(\bm\varepsilon)} = \int d\bm{r}~\phi_p^*(\bm{r}) (\bm\varepsilon \cdot \bm r) \phi_q(\bm{r}).
\label{eq:dpq}
\end{equation}
As shown in~\cref{fig:rixs_states}, intermediate states are high-energy core excitations which are weakly mixed with the lower-energy valence excitations $\ket{E_f}$, that correspond to core-filled electronic configurations. This allows us to simplify the dipole operator as follows
\begin{equation} \label{eq:dip_cvs}
    \hat{D}_{\bm\varepsilon} \approx \sum_{c, v} \sum_{\sigma \in\{\uparrow, \downarrow\}} d^{(\bm\varepsilon)}_{cv} \hat c^\dagger_{v\sigma} \hat c_{c\sigma} + \rm h.c.,
\end{equation}
where $c$ and $v$ denote the indices of core and valence (active) orbitals, respectively, and h.c. stands for the Hermitian conjugate.

This completes the definition of the key observables and operators needed to compute the RIXS cross-section using our proposed quantum algorithm. Before we describe the algorithm, we briefly comment on the challenges of simulating RIXS using classical methods. 

\subsection{Limitations of classical methods for simulating RIXS}
\label{Ssec:class_methods}
The theoretical simulation of RIXS in transition metal complexes stands as one of the most demanding tasks in computational chemistry. Unlike linear absorption spectroscopies, which probe the manifold of excited states via a single photon-matter interaction, RIXS is a coherent photon-in/photon-out scattering process in which an initial excitation populates high-energy core-hole intermediate states, subsequently undergoing radiative decay into valence-excited final states. The challenge of modelling this second-order optical process stems from a combination of factors, including quantum-mechanical interference among the decay channels, vibronic effects in the intermediate states, and the complexity of their electronic structure. The latter is particularly relevant for transition-metal complexes, where the partially filled $3d$ valence shell gives rise to strong static correlation in the ground and low-energy electronic states, causing the breakdown of single-reference methods commonly used in computational X-ray spectroscopy, such as EOM-CC~\cite{krylov_equation--motion_2008} or ADC~\cite{dreuw_algebraic_2015}. Furthermore, single-reference description is often inappropriate for intermediate states which exhibit a high degree of near-degeneracy due to multiplet effects~\cite{neese_advanced_2007}, especially in L-edge RIXS~\cite{josefsson_ab_2012}.

This leaves multi-reference approaches as the only framework suitable for a highly accurate description of the scattering process.  
Among these, the restricted active space self-consistent field (RASSCF) method, combined with second-order perturbation theory (RASPT2)~\cite{casanova_restricted_2022, pinjari_restricted_2014}, has established itself as the de facto standard for simulating RIXS spectra of transition-metal complexes~\cite{bokarev_theoretical_2020,neese_advanced_2007}. RASSCF relies on the active space approximation in which the target wave functions are expanded in the basis of Slater determinants selected on the basis of orbital occupation criteria. This is achieved by designating a subset of orbitals with a fixed total population as active, while keeping the remainder either doubly occupied or vacant in all the determinants. For efficiency, the active space is typically restricted to the orbitals giving rise to static correlation effects, while excitations outside the active space, accounting for dynamic correlation, are treated perturbatively with RASPT2. The choice of active orbitals is therefore critical to the accuracy of RASSCF/RASPT2 calculations and to their computational cost, which scales exponentially with the active space size. 

In RIXS simulations, all participating electronic states need to be treated on equal footing: doing so in practice often requires active spaces that are beyond the reach of available solvers and classical hardware. Failure to incorporate necessary orbitals may result in errors in peak positions and incorrect spectral envelopes~\cite{engel_chemical_2014}. In addition, one must compute a number of high-energy core-excited states (often hundreds or thousands~\cite{delcey2019efficient,delcey_soft_2022}) to ensure convergence of the Kramers-Heisenberg transition amplitude (\cref{eq:rixs_amplitude}). These computational bottlenecks limit the applicability of RASPT2 to active spaces with less than thirteen orbitals~\cite{norell_fingerprints_2018,pinjari_cost_2016,kallman_simulations_2020}, falling short of describing transition metal clusters relevant to the study of oxygen redox. While more recent techniques such as density matrix renormalization group (DMRG) methods hold promise to overcome these limitations, the applicability of dynamical DMRG to RIXS modelling has only been established for relatively small active spaces~\cite{lee2023ab}, and the performance of the TD-DMRG methods~\cite{ren_time-dependent_2022} for the same task is yet to be analyzed.

Recognizing severe limitations of the existing classical simulation techniques, here we propose a fault-tolerant quantum algorithm for evaluating the RIXS cross-section, capable of overcoming the bottlenecks due to the active space size and high density of the core excited states.

\section{Quantum algorithm}
\label{sec:algo}

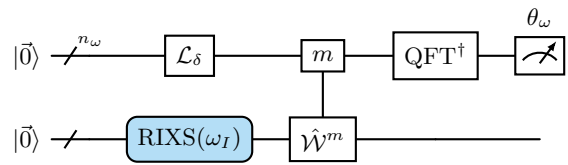
\begin{figure}
    \centering
    \begin{quantikz}
        \lstick{$\ket{\vec 0}$} & \qwbundle{n_\omega} & \gate{\mathcal L_{\delta}} & \gate{m} \vqw{1} & \gate{\rm QFT^\dagger} & \meter{\theta_\omega} \\
        \lstick{$\ket{\vec 0}$} & \qwbundle{} & \gate[style={fill=X3!40,rounded corners}]{\textrm{RIXS}(\omega_I)} & \gate{\hat{\mathcal{W}}^m} &&
    \end{quantikz}
    \caption{Circuit for simulating the RIXS spectrum using walk-based QPE with a qubitized walk operator $\hat{\mathcal{W}}$ \cite{qubitization,qrom,loaiza_mtd}. The blue block indicates the circuit for preparing the initial RIXS state shown in \cref{fig:aa}. The $\mathcal{L}_\delta$ gate prepares a Kaiser lineshape \cite{vs_qsvt}, which minimizes the error coming from the finite precision in QPE \cite{gqpe,optimum_qpe,kaiser_filtering}.}
    \label{fig:qpe}
\end{figure}

The algorithm for computing the RIXS cross-section in Eq.~\eqref{eq:rixs_prob_amplitude} proceeds in two stages: the preparaion of the RIXS state proportional to $\hat R_{\bm\varepsilon_I,\bm\varepsilon_s}(\omega_I)\ket{E_0}$, followed by the application of walk-based QPE, as illustrated in \cref{fig:qpe}. We organize the discussion of the algorithm in four parts. First, we explain the implementation of the quantum walk operator, a core subroutine for both state preparation and QPE. Second, we describe the quantum algorithm for preparing the RIXS state. Third, we describe the subsequent QPE sampling protocol. Finally, we perform the resource requirements analysis for the algorithm, specifically the qubit and Toffoli gate counts.

\subsection{Qubitized walk operator}
We block-encode the Hamiltonian using the block-invariant symmetry shift (BLISS) \cite{bliss,bliss2} technique with the tensor hypercontraction (THC) \cite{thc} factorization, as introduced in Ref. \cite{bliss_thc}. To understand how this technique works, we first consider the BLISS Hamiltonian
\begin{align}
    \hat H_{\rm B}&(\bm\alpha,\bm{\beta}) = \hat H - \alpha_1\hat N_e-\alpha_2\hat N_e^2 \nonumber \\
    &\ \ \ \ \ 
    -\frac{1}{2}\sum_{pq,\sigma}\beta_{pq}\left(\hat c^\dagger_{p\sigma}\hat c_{q\sigma}(\hat N_e-N_e) + \textrm{h.c.}\right) \\
    &\equiv \sum_{pq,\sigma}\tilde h_{pq}(\bm\alpha,\bm\beta) \hat c_{p\sigma}^\dagger \hat c_{q\sigma} \nonumber \\
    &\ \ \ \ \ +\frac{1}{2}\sum_{pqrs,\sigma\sigma'} \tilde V_{pqrs}(\bm\alpha,\bm\beta) \hat c_{p\sigma}^\dagger \hat c_{q\sigma} \hat c_{r\sigma'}^\dagger \hat c_{s\sigma'} 
\end{align}
where $\hat{H}$ is the electronic Hamiltonian in~\cref{eq:hamiltonian}, $\hat N_e=\sum_{p\sigma} \hat c^\dagger_{p\sigma}\hat c_{p\sigma}$ is the total particle number operator and $N_e$ is the number of electrons of the ground-state $\hat N_e\ket{E_0}=N_e\ket{E_0}$. The $\bm\alpha$ and $\bm\beta$ parameters here constitute free parameters that can be optimized as to minimize the Hamiltonian's 1-norm. Here we have skipped the constant term for notational clarity. Note that the expectation value of any function of this Hamiltonian will be identical as that of the original Hamiltonian as long as the wavefunction corresponds to a system with $N_e$ electrons. Both Jordan-Wigner and Bravyi-Kitaev mappings transform the particle number operator as $\hat c_{p\sigma}^\dagger\hat c_{p\sigma} \rightarrow (\hat z_{p\sigma}+\hat 1)/2$, where $\hat z_{p\sigma}$ is the Pauli $Z$ operator acting on the qubit associated with spin-orbital $p\sigma$ \cite{jw_1,jw_2,bk_1,bk2}. The tensor hypercontraction framework factorizes this BLISS Hamiltonian as
\begin{align}
    \hat H_{B}(\bm{\alpha},\bm{\beta},&\bm{\zeta},\hat{\bm{U}}) = -\frac{1}{2}\sum_{p,\sigma} t_p \hat U_0^\dagger \hat z_{p\sigma} \hat U_0 \nonumber \\
    &\ \ \ \ \ + \frac{1}{8}\sum_{\mu\nu=1}^{N_T}\sum_{\sigma\sigma'} \zeta_{\mu\nu} \hat U_\mu^\dagger \hat z_{1\sigma}\hat U_\mu\hat U_\nu^\dagger \hat z_{1\sigma'}\hat U_\nu,
\end{align}
where the tensor $\zeta_{\mu\nu}$ and the real orbital rotations $\hat U$'s correspond to the parameters of the THC decomposition, alongside the associated rank $N_T$. Note that the values of $t_p$ are obtained by diagonalizing the associated one-body matrix $\bm \kappa$ with elements
\begin{equation}
    \kappa_{pq} = \tilde h_{pq} - \frac{1}{2}\sum_r V_{prrq} + \sum_r \tilde V_{pqrr}-\alpha_1\delta_{pq}+2N_e\beta_{pq}
\end{equation}
such that
\begin{equation}
    \sum_{pq,\sigma} \kappa_{pq} \hat c_{p\sigma}^\dagger \hat c_{q\sigma} = \sum_{p,\sigma} t_p \hat U_0^\dagger \hat c_{p\sigma}^\dagger \hat c_{p\sigma} \hat U_0,
\end{equation}
where we here skip explicit dependence to the BLISS parameters for clarity. This decomposition of the Hamiltonian as a linear combination of unitaries has the associated 1-norm
\begin{equation}
    \lambda(\bm{\alpha},\bm\beta,\bm{\zeta}) = \sum_p |t_p(\bm\alpha,\bm\beta)| + \frac{1}{2}\sum_{\mu\nu} |\zeta_{\mu\nu}|-\frac{1}{4}\sum_\mu |\zeta_{\mu\mu}|.
\end{equation}
The BLISS-THC decomposition is then obtained by minimizing the cost function
\begin{align}
    C(\bm{\alpha},\bm{\beta},\bm{\zeta},\hat{\bm{U}}) &\equiv \frac{1}{2} \sum_{pqrs}\left[\tilde V_{pqrs}(\bm\alpha,\bm{\beta})-\sum_{\mu\nu} \zeta_{\mu\nu}\chi^{\mu\nu}_{pqrs}\right]^2 \nonumber \\
    &\ \ \ \ \ + \rho\cdot\lambda(\bm \alpha,\bm\beta,\bm{\zeta}),
\end{align}
where we have defined the rotation tensor $\chi_{pqrs}^{\mu\nu} = U^{(\mu)}_{1p} U^{(\mu)}_{1q} U^{(\nu)}_{1r} U^{(\nu)}_{1s}$ for $U_{pq}^{(\mu)}$, the matrix element defining the action of the orbital rotation $\hat U_\mu$ such that $\hat U_\mu^\dagger \hat c^\dagger_p c_q\hat U_\mu = \sum_{rs} U^{(\mu)}_{pr} U^{(\mu)}_{qs} \hat c^\dagger_r\hat c_s$. Here $\rho$ acts as the regularization parameter controlling the strength of the 1-norm penalty, which has been shown to help keep the 1-norm small during the cost function minimization procedure \cite{bliss_thc}.

Once the BLISS-THC decomposition of the Hamiltonian has been obtained, the associated PREPARE and SELECT oracles can be constructed for block-encoding $\hat H/\lambda$ and constructing the qubitized walk operator 
\begin{equation}
    \hat{\mathcal{W}} = \hat{\mathcal{R}}\cdot \mathtt{PREP\cdot SEL\cdot PREP}^\dagger,
\end{equation}
which acts as $e^{\pm i \arccos \hat H/\lambda}$ \cite{qubitization}. Here $\hat{\mathcal{R}}=(\hat I-2\ket{0}\bra{0})\otimes\hat I$ is the reflection on the auxiliary qubits coming from PREPARE. On a high level, the PREPARE oracle will load the $\zeta_{\mu\nu}$ coefficients as $\sqrt{\zeta_{\mu\nu}}\ket{\mu,\nu}$, while the SELECT oracle consists of Pauli $Z$ gates that are conjugated by the orbital rotations $\hat U_\mu$; these orbital rotations are implemented using $N_a-1$ Givens rotations where the associated angles are loaded by using a QROM on the $\ket{\mu}$ and $\ket{\nu}$ registers \cite{df}. The full details for how these oracles are implemented are presented in Refs. \cite{thc,bliss_thc}.

\subsection{Preparation of the RIXS state}
We now discuss the preparation of the RIXS state
\begin{equation} \label{eq:rixs_state}
    \ket{R_{\bm{\varepsilon}_I, \bm{\varepsilon}_S}(\omega_I)} \equiv \frac{\hat R_{\bm{\varepsilon}_I, \bm{\varepsilon}_S}(\omega_I) \ket{E_0}}{|R_{\bm\varepsilon_I,\bm\varepsilon_S}(\omega_I)|},
\end{equation}
where we have defined the normalization constant
\begin{equation}
    |R_{\bm\varepsilon_I,\bm\varepsilon_S}(\omega_I)| \equiv \sqrt{\|\hat R_{\bm{\varepsilon}_I, \bm{\varepsilon}_S}(\omega_I) \ket{E_0}\|^2}.
\end{equation}
The RIXS state is prepared by performing a block-encoding of the $\hat R_{\bm{\varepsilon}_I,\bm{\varepsilon}_S}(\omega_I)$ operator (see~\cref{eq:rixs_operator}), which is used inside an amplitude amplification framework to ensure its successful preparation. Now we discuss our block-encoding for this operator.

First we prepare the dipole-perturbed state
\begin{equation} \label{eq:dipole_state}
    \ket{D_{\bm\varepsilon_I}} \equiv \frac{\hat D_{\bm{\varepsilon}_I}\ket{E_0}}{|D_{\bm\varepsilon_I}|},
\end{equation}
where we have defined the normalization constant
\begin{equation}
    |D_{\bm\varepsilon_I}|\equiv \sqrt{\left\|\hat D_{\bm{\varepsilon}_I}\ket{E_0}\right\|^2},
\end{equation}
and assumed that we have access to the electronic ground-state. This means that the state $\ket{D_{\bm\varepsilon_I}}$ will also be classically accessible, and can be directly loaded on the quantum computer using different approaches, such as tensor-based representations \cite{tensor_prep_1,tensor_prep_2} or as a sum of Slater determinants \cite{sum_of_slaters}. We defined $\hat U_{\bm\varepsilon_I}$ as the associated unitary that transforms the computational $\ket{\vec 0}$ state into $\ket{D_{\bm\varepsilon_I}}$. Here we use the simplified dipole operator from Eq.\eqref{eq:dip_cvs}, noting that only intermediate states with core electron excitations will have a contribution to the RIXS spectrum.

The next step is to apply the Green's function, for which we will use a block-encoding of the Hamiltonian $\hat H$ in conjunction with the generalized quantum signal processing (GQSP) framework. GQSP allows to block-encode any polynomial of a unitary, namely $P(\hat U)=\sum_{n=-k}^{m} a_n \hat U^n$ for $a_n\in\mathbb C$, using $m$ controlled applications of $\hat U$, $k$ controlled applications of $\hat U^\dagger$, and $m+k+1$ single qubit $\rm SU(2)$ rotations \cite{gqsp,dyn_cooling} if the condition $|P(e^{i\theta})|^2\leq 1$ for all $\theta\in\mathbb R$ is satisfied. We expand the Green's function $\hat G(\omega_I,\Gamma)$ using a $K_G$th degree Chebyshev expansion
\begin{align}
    \Gamma\hat G(\omega_I,\Gamma) &= \frac{\Gamma}{\omega_I-(\hat H-E_0)+i\Gamma} \\
    &\approx \sum_{k=0}^{K_G} c_k(\omega_I,\Gamma,\lambda)\cdot T_k\left(\frac{\hat H}{\lambda}\right). \label{eq:cheby_green}
\end{align}
The rescaling by $\Gamma$ is done to guarantee that the spectral norm $|\Gamma\hat G(\omega_I,\Gamma)|\leq 1$, which allows us to implement the associated polynomial using GQSP by using $K_G$ controlled applications of the qubitized walk operator $\hat{\mathcal W}$, $K_G$ controlled applications of $\hat{\mathcal{W}}^\dagger$, plus $2K_G+1$ general $\rm SU(2)$ single qubit rotations that can be realized by a set of $4K_G+2$ $R_z$ and $2K_G+1$ $R_y$ rotations. Refs. \cite{gqsp,gqsp_angles} present a discussion of how to efficiently find the rotation angles. A more in-depth discussion of the Chebyshev expansion for the Green's function alongside its GQSP implementation is provided in \cref{app:circuits}. Note that the walk operator is associated with the rescaled Hamiltonian $\hat H/\lambda$, which effectively rescales the x-axis by a $1/\lambda$ factor. In turn, this rescaling causes all features to sharpen, as the first-order derivative is multiplied by $\lambda$. This reduction in smoothness results in a required polynomial degree $K_G$ that scales as $\tilde{\mathcal{O}}(\lambda)$.

The final step for preparing the RIXS state requires the application of the dipole operator $\hat D^\dagger_{\bm\varepsilon_S}$ [\cref{eq:dip_cvs}], which can be expressed as an orbital rotation conjugating a linear combination of Pauli Z operators and implemented with an optimal 1-norm (see~\cref{app:circuits}).

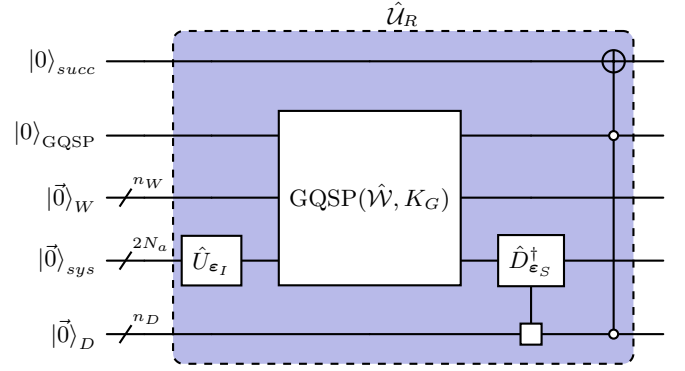
\begin{figure}[!t!]
    \centering
    \resizebox{0.5\textwidth}{!}{
    \begin{quantikz}
        \lstick{$\ket{0}_{succ}$} && \gategroup[5,steps=4,style={dashed,rounded
			corners,fill=X1!40, inner
			xsep=0pt},background,label style={label
			position=above,anchor=north,yshift=0.3cm}]{$\hat{\mathcal{U}}_R$} &&& \targ{} & \\
        \lstick{$\ket{0}_{\rm GQSP}$} &&& \gate[3]{\textrm{GQSP}(\hat{\mathcal{W}},K_G)} && \octrl{0} &  \\
        \lstick{$\ket{\vec 0}_W$} & \qwbundle{n_W} &&&&& \\
        \lstick{$\ket{\vec 0}_{sys}$} & \qwbundle{2N_a} & \gate{\hat U_{\bm\varepsilon_I}} && \gate{\hat D^\dagger_{\bm\varepsilon_S}} && \\
        \lstick{$\ket{\vec 0}_D$} & \qwbundle{n_D} &&& \gate{}\wire[u][1]{q} & \octrl{-4} &
    \end{quantikz}}
    \caption{Circuit for the block-encoding $\hat{\mathcal{U}}_R$ [Eq.\eqref{eq:U_R}] to prepare the RIXS state [Eq.\eqref{eq:rixs_state}] flagged by $\ket{1}_{succ}$. The unitary $\hat U_{\bm\varepsilon_I}$ prepares the dipole-perturbed state $\ket{D_{\bm\varepsilon_I}}$ [Eq.\eqref{eq:dipole_state}]. $\textrm{GQSP}(\hat{\mathcal{W}},-K_G,K_G)$ implements a $K_G$th degree approximation to the Green's function [Eq.\eqref{eq:cheby_green}] via the walk operator $\hat{\mathcal{W}}$ using generalized quantum signal processing, followed by a block-encoding of the dipole operator  $\hat D^\dagger_{\bm\varepsilon_S}$. The $\ket{\cdot}_W$ register denotes ancillas used by the walk operator, while $\ket{\cdot}_D$ are the dipole block-encoding qubits, $\ket{\cdot}_{sys}$ are the qubits encoding the system's wavefunction, and $\ket{\cdot}_{\rm GQSP}$ is the additional qubit for GQSP rotations.}
    \label{fig:state}
\end{figure}

At this stage, we have provided all the techniques to block-encode the operator $\hat R_{\bm\varepsilon_I,\bm\varepsilon_S}(\omega_I)$, which acts as
\begin{equation} \label{eq:U_R}
\hat{\mathcal{U}}_R  \equiv
    \begin{bmatrix}
        \frac{\Gamma}{\lambda_{D}^{(\bm\varepsilon_S)}}\hat D^\dagger_{\bm\varepsilon_S}\hat G(\omega_I,\Gamma)\hat U_{\bm\varepsilon_I} & \cdot \\
        \cdot & \cdot
    \end{bmatrix}
\end{equation}
as shown in \cref{fig:state}. Note that we have used one additional qubit to flag whether the block-encoding was successful.

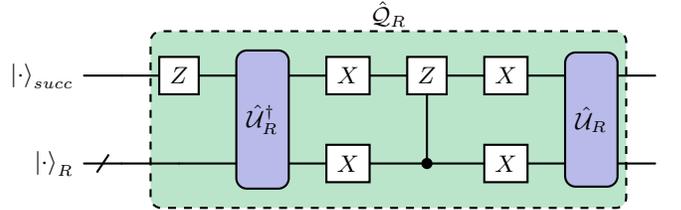
\begin{figure}
    \centering
    \begin{quantikz}
        \lstick{$\ket{\cdot}_{succ}$} && \gate{Z} \gategroup[2,steps=6,style={dashed,rounded
			corners,fill=X2!40, inner
			xsep=0pt},background,label style={label
			position=above,anchor=north,yshift=0.3cm}]{$\hat{\mathcal{Q}}_R$} & \gate[2,style={fill=X1!40,rounded corners}]{\hat{\mathcal{U}}_R^\dagger} & \gate{X} & \gate{Z} & \gate{X} & \gate[2,style={fill=X1!40,rounded corners}]{\hat{\mathcal{U}}_R} & \\
        \lstick{$\ket{\cdot}_R$} & \qwbundle{} &&& \gate{X} & \ctrl{-1} & \gate{X} &&
    \end{quantikz}
    \caption{Construction of Grover iterate $\hat{\mathcal{Q}}_R$ for amplitude estimation and amplification on block-encoding $\hat{\mathcal{U}}_R$ (\cref{fig:state}). The register $\ket{\cdot}_R$ collects all system and ancilla qubits (walk, GQSP, and dipole) excluding the success register, with the associated $X$ gate acting on all its qubits.}
    \label{fig:grover}
\end{figure}

Next, we use amplitude amplification on the block-encoding $\hat{\mathcal{U}}_R$ to boost its success probability close to 1. We start by considering its success probability, which is given by
\begin{equation} \label{eq:succ}
    P_R \equiv \left(\frac{\Gamma|R_{\bm\varepsilon_I,\bm\varepsilon_S}(\omega_I)|}{\lambda_D^{(\bm\varepsilon_S)} |D_{\bm\varepsilon_I}|}\right)^2.
\end{equation}
Although fixed point amplitude amplification \cite{fp_qae_1,fp_qae_2} could be used without prior knowledge of $P_R$, in this case it is advantageous to first determine $P_R$ and then use ``textbook'' amplitude amplification \cite{grover,brassard2000quantum} which has better prefactors. $P_R$ can be determined most efficiently using amplitude estimation on the Grover iterate that is built from the block-encoding $\hat{\mathcal{U}}_R$ \cite{brassard2000quantum, iqae, qsp_qae}. The construction of this Grover iterate is shown in \cref{fig:grover}. Note that the overall cost of estimating $P_R$ is subdominant compared to the cost of the full QPE circuit repetitions.

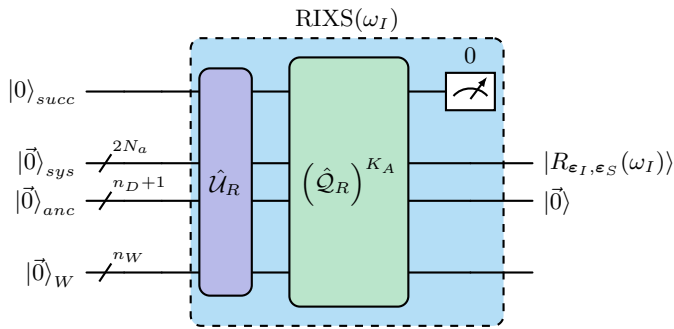
\begin{figure}
    \centering
    \begin{quantikz}
        \lstick{$\ket{0}_{succ}$} &&& \gate[4,style={fill=X1!40,rounded corners}]{\hat{\mathcal{U}}_R} \gategroup[4,steps=3,style={dashed,rounded
			corners,fill=X3!40, inner
			xsep=0pt},background,label style={label
			position=above,anchor=north,yshift=0.3cm}]{$\textrm{RIXS}(\omega_I)$} & \gate[4,style={fill=X2!40,rounded corners}]{ \left(\hat{\mathcal{Q}}_R\right)^{K_{A}}} & \meter{0} \\
        \lstick{$\ket{\vec 0}_{sys}$} & \qwbundle{2N_a} &&&&& \rstick{$\rixs$} \\
        \lstick{$\ket{\vec{0}}_{anc}$} & \qwbundle{n_D+1} &&&&& \rstick{$\ket{\vec 0}$} \\
        \lstick{$\ket{\vec{0}}_{W}$} & \qwbundle{n_W} &&&&&
    \end{quantikz}
    \caption{Circuit for preparing RIXS state with high success probability using amplitude amplification on block-encoding $\hat{\mathcal{U}}_R$ (\cref{fig:state}) with associated Grover iterate $\hat{\mathcal{Q}}_R$ (\cref{fig:grover}). Ancilla registers here are dipole and GQSP qubits.}
    \label{fig:aa}
\end{figure}

Once an estimate for $P_R$ has been found, the associated number of rounds of amplitude amplification
\begin{equation} \label{eq:ka}
    K_{A} = \floor{\frac{\pi}{4\arcsin\sqrt{P_R}}},
\end{equation}
can be used to boost the success probability close to 1. The circuit for building the RIXS state with high success probability is shown in \cref{fig:aa}, requiring a total of $2K_{A}+1$ calls to $\hat{\mathcal{U}}_R$. \\

\subsection{Walk-based QPE}
Once the RIXS state is prepared, we can use it as the initial state of QPE to directly obtain the RIXS spectrum for a given frequency $\omega_I$ of the incident X-ray photon. While two-dimensional RIXS maps could also be accessed using  generalized QPE \cite{gqpe}, here we focus on the quantum simulation of high-resolution RIXS spectra for selected values of $\omega_I$ (see~\cref{ssec:class_sim}), in line with the experimental requirements.

The application of the QPE circuit applying the time-evolution operator $e^{-i\hat H t}$ on some initial state $\ket{\psi}=\sum_n c_n \ket{E_n}$ yields, up to discretization and finite precision considerations, a frequency $\omega$ that is sampled from the distribution
\begin{equation}
    P_t\big(\omega;\ket{\psi}\big) = \sum_n |c_n|^2 \delta(\omega - E_n).
\end{equation}
In practice, errors coming from discretization and finite precision are most efficiently reduced by preparing the QPE register in a special state, rather than the usual uniform superposition achieved through Hadamard gates on all qubits \cite{qrom}. As can be seen through the convolution theorem, this modification has the effect of replacing the infinite precision Dirac delta of QPE  by some finite precision lineshape \cite{gqpe}: specifically, we opt for using the Kaiser lineshape, as it has been shown to yield optimal convergence properties \cite{vs_qsvt,kaiser_filtering}. Noting that the walk operator $\hat{\mathcal{W}}$ shares eigenstates $\ket{E_n}$ with $\hat H$ with associated eigenvalues $\arccos E_n/\lambda$, it follows that by performing QPE using the walk operator instead of the time-evolution will then sample values $\theta_\omega$ from the distribution
\begin{equation} \label{eq:qpe_prob}
    P_{\mathcal{W}}\big(\theta_\omega;\ket{\psi}\big) = \sum_n |c_n|^2 \delta\left(\theta_\omega - \arccos\frac{E_n}{\lambda}\right).
\end{equation}
Writing the RIXS state as $\rixs = \sum_f W_{f0}(\omega_I,\bm\varepsilon_I,\bm\varepsilon_S)\ket{E_f}$, it becomes apparent how application of the walk-based QPE circuit shown in \cref{fig:qpe} can be used to recover the RIXS amplitude in Eq.\eqref{eq:rixs_prob_amplitude} by transforming the associated x-axis as $\omega=E_0+\lambda\cos\theta_\omega$, which is summarized by the following expression:
\begin{equation} \label{eq:qpe_prob}
    P_{\bm\varepsilon_I,\bm\varepsilon_S}(\omega_I,\omega) \propto P_{\mathcal{W}}\big(\lambda\cos\theta_{\omega};\rixs\big).
\end{equation}
The number of calls to the walk operator in each QPE circuit for obtaining a target accuracy $\epsilon_\omega$ for the energy loss $\omega$ can be calculated as 
\begin{equation} \label{eq:walk_calls}
    N(\epsilon_\omega) = \ceil{\frac{\pi\lambda}{\sqrt{2}\epsilon_\omega}},
\end{equation}
requiring $n_\omega=\ceil{\log_2 N(\epsilon_\omega)}$ ancilla qubits for the phase estimation register \cite{qrom}.

\subsection{Resource requirements}
We now discuss the associated resource requirements for simulating the RIXS cross-section. We start by providing a top-to-bottom summary for obtaining the cost of our algorithm:
\begin{itemize}
    \item Walk-based QPE (\cref{fig:qpe}): $2^{n_\omega}$ calls to the walk operator $\hat{\mathcal{W}}$, and one RIXS state preparation. Toffoli costs of preparing the Kaiser lineshape and performing the quantum Fourier transform are negligible. This routine requires $n_W$ ancilla qubits corresponding to those used by the walk operator, alongside $n_\omega$ qubits for QPE and $2N_a$ for representing the electronic system.
    \item RIXS state preparation (\cref{fig:aa}): $2K_A+1$ calls to the block-encoding $\hat{\mathcal{U}}_R$. Toffoli cost of non-$\hat{\mathcal{U}}_R$ components appearing in the Grover iterate (\cref{fig:grover}) is negligible. For the qubit count besides those used by $\hat{\mathcal{U}}_R$, the multicontrolled $Z$ appearing in the Grover iterate can be done using 2 additional ancilla qubits \cite{multicontrol}. Note that this construction for multicontrolled $Z$ requires about 4 times more Toffolis than the one using temporary ANDs. However, since the Toffoli cost of this routine is insignificant when compared to $\hat{\mathcal{U}}_R$, we here chose a strategy that avoids duplicating the required number of qubits.
    \item Block-encoding $\hat{\mathcal{U}}_R$ (\cref{fig:state}): There are $K_G$ calls to $\hat{\mathcal{W}}$ and $K_G$ calls to $\hat{\mathcal{W}}^\dagger$ from GQSP; all other Toffoli costs in this routine are negligible in comparison. This routine requires $n_D=3\ceil{\log_2 N_a}+2\aleph_\mu+2$ qubits for block-encoding the dipole operator (see \cref{app:circuits} for more details), $n_W$ qubits from the walk operator implementation, one additional qubit for GQSP, and one additional qubit for the success register. Here we have considered $\aleph_\mu$ qubits associated with the precision used for loading the dipole coefficients. Note that implementation of the multicontrolled NOT operation in this routine can be done using the same 2 clean ancillas for the multicontrolled $Z$ in the Grover iterate for no additional ancilla cost. The total number of additional ancillas used by this routine is then $n_D+n_W+2$. The qubits from the walk operator $n_W$ here are shared with those used for the walk-based QPE routine, while the resource state used for the phase gradient technique for implementing the Givens rotations in the walk operator can also be used for doing the orbital rotation for the block-encoding of the dipole operator.
    \item  Walk operator $\hat{\mathcal{W}}$ using BLISS-THC decomposition \cite{bliss_thc}: two calls to the PREPARE oracle and one to the SELECT oracle associated with the THC factorization \cite{bliss_thc,thc}. The cost of these routines will be determined by: the alias sampling precision $\aleph$, the Givens rotation precision $\beth$, the THC rank $N_T$, and the number of batches for loading Givens rotation angles. In this work we consider the version without any batching, which minimizes the gate cost in exchange for using more qubits. We use the notation of $\mathcal T_{W}(\aleph,\beth,N_T)$ and $n_W(\aleph,\beth,N_T)$ for the Toffoli gates and ancilla qubits used by $\hat{\mathcal{W}}$ respectively.
\end{itemize}

Finally, we note how the $n_D+1$ qubits used for block-encoding the dipole operator and GQSP, alongside the $2$ qubits for the multicontrolled NOT and $Z$, and the success qubit for amplitude amplification, are all returned to the zero state after successful application of the RIXS state preparation in \cref{fig:state}, so that those qubits can then be used for the QPE algorithm. The total number of qubits required by the algorithm then becomes
\begin{equation}
    n_{tot} = 2N_a+\textrm{max}\{n_\omega;n_D+4\}+n_W (\aleph,\beth,N_T)
\end{equation}
with the number of Toffoli gates being
\begin{equation}
   \mathcal T_{tot} = \big((2K_A+1) \cdot 2K_G + 2^{n_\omega}\big)\cdot \mathcal T_W(\aleph,\beth,N_T)
\end{equation}
up to subdominant factors. Note that the Toffoli cost will have a $\tilde{\mathcal{O}}(\lambda)$ scaling with the Hamiltonian 1-norm through $2^{n_\omega}$ and $K_G$.

\section{Application}
\label{sec:aplication}
\begin{figure}[t]
    \centering
    \includegraphics[width=0.7\linewidth]{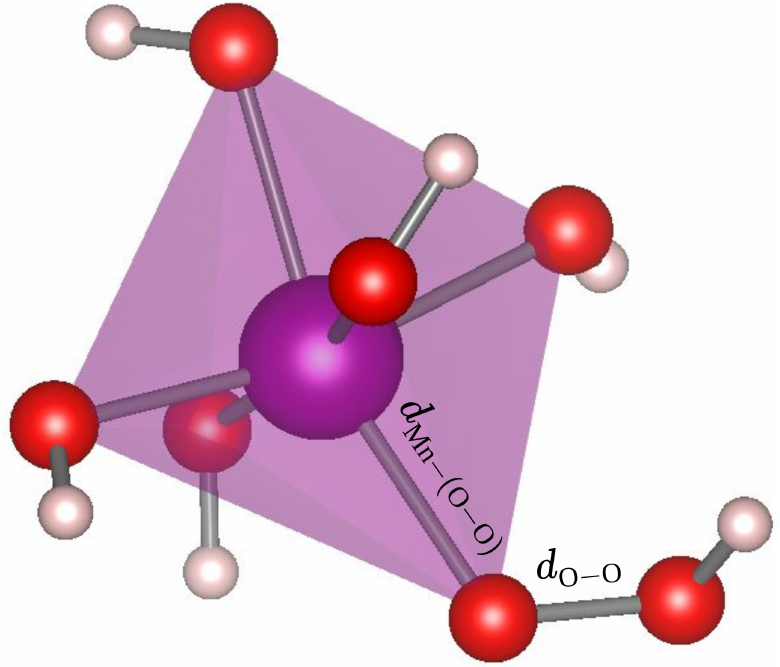}
    \caption{Atomic structure of the $\text{MnO}_7\text{H}_6$ molecular cluster used to perform resource estimation for quantum simulation of the RIXS spectra. The oxygen dimer (O-O), with bond length $d_{\text{O}-\text{O}}=1.27~\text{\AA}$, is bonded to the manganese species, which is coordinated by oxygen species of the cathode material. The distance between the oxygen atom in the O-O dimer and the transition metal is $d_{\text{Mn}-(\text{O-O})} = 2.09~\text{\AA}$.}
    \label{fig:cluster}
\end{figure}

In this section, we apply the quantum algorithm we proposed to a prototypical molecular cluster predicted to form in Li-excess cathodes~\cite{gao2025clarifying}. \cref{fig:cluster} shows the atomic structure of the cluster ($\text{MnO}_7\text{H}_6$) used to estimate the cost of the quantum algorithm developed in \cref{sec:algo}. Its structure was cut out of the relaxed delithiated Li-rich cathode supercell reported in Ref.~\cite{gao2025clarifying}. We use hydrogen atoms to passivate dangling bonds. The cluster consists of an oxygen dimer (O-O) bonded to a manganese atom, which is coordinated with oxygen species of the cathode material.

\subsection{Construction of the active spaces}
\label{ssec:system}

To build the second-quantized observables defined in~\cref{ssec:theory}, we need to define the set of molecular orbitals for our chosen cluster. We performed a ROHF calculation in 6-31$\text{G}^{**}$ Gaussian basis assuming total spin projection $S_{z} = 1/2$ using the code \texttt{PySCF}~\cite{sun_pyscf_2018}. Once a stable solution was obtained, we generated a series of active spaces to define effective Hamiltonians for resource estimation. To ensure a chemically motivated selection of the active orbitals, we applied the Atomic Valence Active Space (AVAS) approach of Sayfutyarova {\it et al.}~\cite{sayfutyarova_automated_2017} with a tunable threshold. In AVAS, the active space is spanned by the orbitals selected among the eigenvectors of a projector on a user-defined space of atomic orbitals $|\chi_i\rangle$ \begin{align}
    P_A = \sum_{ij} |\chi_i\rangle (\mathbf{S}^{-1})_{ij} \langle\chi_j|\label{eq:avas_proj},
\end{align}
where $\mathbf{S}_{ij} = \langle\chi_j|\chi_i\rangle$ is an overlap matrix of reference orbitals defined in a suitable basis.
\begin{figure*}[t]
\centering
    \includegraphics[width=1\linewidth]{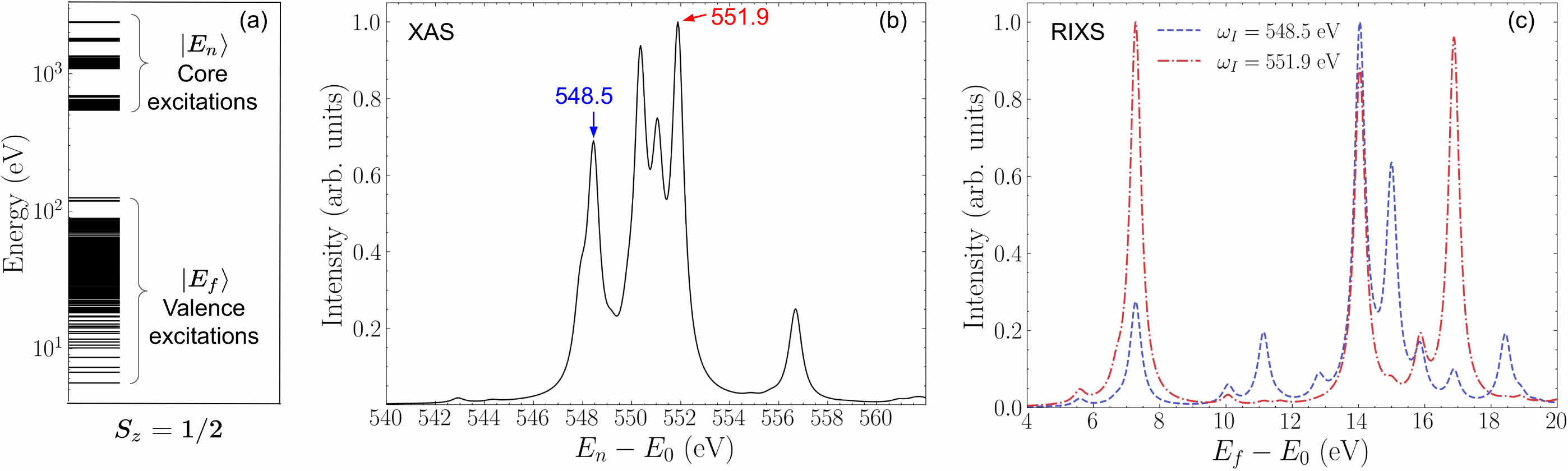}
    \caption{Classical simulations of (a) the spectrum of electronic states, and (b) the X-ray absorption (XAS) and (c) resonant inelastic X-ray scattering (RIXS) spectra of the $\text{MnO}_7\text{H}_6$ molecular cluster shown in~\cref{fig:cluster}. The electronic states with spin projection $S_z=1/2$ were computed within full configuration interaction (FCI) method using an active space with $N_a=8$ orbitals and $N_e=11$ electrons. The energies of the ground, intermediate, and final excited  states involved in the scattering process (see~\cref{eq:rixs_amplitude_simp}) are denoted as $E_0$, $E_n$ and $E_f$, respectively. The RIXS spectra were simulated at two incident photon frequencies that correspond to resonant features in the XAS spectrum.}
    \label{fig:xas-rixs}
\end{figure*}
The active orbitals are obtained by diagonalizing $P_A$ on occupied and virtual molecular orbital subspaces separately, excluding the eigenvectors with small eigenvalues below threshold from the active space. The size of an active space can therefore be systematically increased by lowering the threshold. 

Here, we applied AVAS in a two-step manner, by first constraining the number of electrons in the active space, then systematically augmenting it with virtual orbitals required to capture static and dynamic correlation in the states involved in the RIXS process. The details of this procedure and the AVAS parameters are presented in~\cref{app:avas}.
Guided by previous studies of the electronic structure of transition metal oxide cathode materials~\cite{okubo_molecular_2017,roychoudhury_deciphering_2021}, we incorporated a baseline selection of O $1s$ and hybridized O $2p$ + Mn $3d$ orbitals centered on the peroxide fragment in all the active spaces, so as to accurately describe the initial O K-edge excitation. This set was further expanded to include non-hybridizing Mn $3d$ orbitals, expected to be important for the accurate description of the low-energy $d$-$d$ and charge-transfer excited states of the cluster, designated as final states in RIXS. Finally, for larger active  spaces (with $N_a \ge 19$) the Mn $4p$ orbitals were incorporated as well. This way, we generated a set of Hamiltonian instances ranging from exactly diagonalizable to completely classically intractable, with Hamiltonian FCI matrix dimensions listed in~\cref{tab:re_rixs}.

\subsection{Classical simulations}
\label{ssec:class_sim}
To build intuition around RIXS spectra of the systems being studied, as well as around the nature of the intermediate and final RIXS states and importantly to fix key parameters needed for resource estimation, we have performed classical simulation of RIXS spectra at the O K-edge. \cref{fig:xas-rixs} illustrates the classical computational pipeline for simulating the RIXS spectrum of our molecular cluster. We use the full configuration interaction (FCI) method to describe the cluster electronic states, and truncated the active space to $N_a=8$ orbitals and $N_e=11$, corresponding to a Hamiltonian matrix with dimension $D_\text{FCI} = 1568$. \cref{fig:xas-rixs}(a) plots the energies of the FCI states with total spin projection $S_z=1/2$. The valence excitations span the energy range of $[5, 119]$ eV, whereas the core excitations, in resonance with soft X-rays, have energies exceeding $543$ eV. The FCI core-excited states were used to simulate the XAS spectrum~\cite{fomichev2025fast} shown in~\cref{fig:xas-rixs}(b). Similarly to what is done experimentally, we chose two resonances in the XAS spectrum ($548.5$ eV and $551.9$ eV) to simulate the RIXS spectra plotted in~\cref{fig:xas-rixs}(c). For a given frequency $\omega_I$, we sum over all intermediate states in the energy window $|(E_n-E_0)-\omega_I|\leq 50$ eV to compute the transition amplitude $|W_{f0}|^2$ (see~\cref{eq:rixs_amplitude}). The energy width of the intermediate states was set to $\Gamma_n=0.3$ eV, and the polarization vectors of the incoming and scattered photons were defined as $\bm\varepsilon_I=\bm\varepsilon_S=(1, 0, 0)$. Finally, the spectra in~\cref{fig:xas-rixs}(c) were built by smoothing the Dirac delta function in~\cref{eq:rixs_cross_section} with a Lorentzian with width $\eta=0.2$ eV to account for the lifetime broadening of the final states.

The calculated RIXS spectra in \cref{fig:xas-rixs}(c) show strong resonance effects, evidenced by significant variations of the peak relative intensities as a function of the incoming X-ray frequency. For instance, X-ray photons  with energy $\omega_I=548.5$ eV primarily excite two peaks centered around $14$ eV and $15$ eV. In contrast, for the higher value of $\omega_I=551.9$ eV three dominant peaks appear at approximately $7$ eV, $14$ eV, and $17$ eV. It can also be noted that certain excited states, specifically those at $11$ eV, $15$ eV, and $18.5$ eV, are only visible for $\omega_I=548.5$ eV. We inspected the wavefunctions of the excited states with the highest RIXS amplitudes and observed that all valence orbitals in the active space have non-negligible contributions to the correlated state. Most states consist of singly-excited configurations involving orbitals in the oxygen dimer mixed with double excitations involving more delocalized orbitals.

\subsection{Resource estimation}
\begin{figure}
    \centering
    \includegraphics[width=0.95\linewidth]{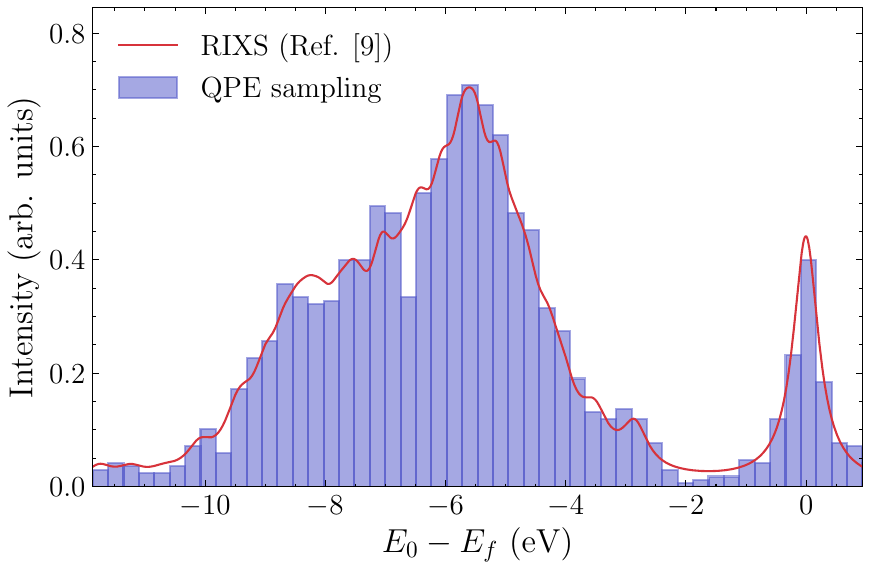}
    \caption{Experimental RIXS spectrum of a charged Li-rich NMC cathode reported in Ref.~\cite{gao2025clarifying} and its QPE sampling reconstruction with 2,000 shots.}
    \label{fig:shots}
\end{figure}

We now discuss all the details for obtaining RIXS spectra of the $\text{MnO}_7\text{H}_6$ cluster using the quantum algorithm in \cref{sec:algo}, and report resource estimates in \cref{tab:re_rixs}. The workflow is analogous to the classical simulation pipeline discussed in~\cref{ssec:class_sim}. First, we determine the frequencies $\omega_I$ from the XAS spectrum for which we want to simulate the RIXS cross section. For a given $\omega_I$, the quantum algorithm in~\cref{fig:qpe} will directly sample the excitation energies $E_f-E_0$ (energy loss) of the cluster final states. The full RIXS spectrum shown in~\cref{fig:xas-rixs} (c) is reconstructed by repeated sampling of the circuit of \cref{fig:qpe}. A spectral reconstruction example is shown in \cref{fig:shots}, where we empirically determine that around $2,000$ circuit repetitions are sufficient for recovering a spectrum at the accuracy demanded by the application studied in this work. This was done by directly drawing random samples with respect to the probability distribution defined by the (normalized) spectrum given in Ref. \cite{gao2025clarifying}, with bin size 0.2 eV as a typical broadening in RIXS experiments.

Based on the~\cref{eq:dip_cvs} for the dipole operator, only core-excited states in the dipole-perturbed state $\ket{D_I}$ [Eq.\eqref{eq:dipole_state}] will have significant contributions to the RIXS spectrum. In addition, we know that the action of the dipole operator $\hat{D}_{\bm\varepsilon_S}$ must transform core-excited $\ket{E_n}$'s into core-filled $\ket{E_f}$'s. This makes both quantities $|D_{\bm\varepsilon_I}|$ and $\lambda_D^{(\bm\varepsilon_S)}$ appearing in the success probability in Eq.\eqref{eq:succ} significantly smaller, which boosts the success probability of the block-encoding operator $\hat{\mathcal{U}}_R$, making the algorithm cheaper by requiring less rounds of amplitude amplification.

We now provide all the hyperparameters that were used for obtaining the resource estimates in \cref{tab:re_rixs} using PennyLane \cite{pennylane}. For the implementation of the walk operator with BLISS-THC, we used the precision parameters $\aleph=\beth=13$ which were shown to yield accurate results for even significantly larger systems \cite{bliss_thc}, with a rank of $N_T=3N_a$. The same number of precision qubits $\aleph_\mu=13$ was used for the coherent alias sampling precision when block-encoding the dipole operator. The formula in Eq.\eqref{eq:walk_calls} was used with the accuracy corresponding to the Lorentzian broadening of final states $\eta=0.2\ \rm eV$, namely $N(\eta)\approx 303\cdot\lambda$, while the broadening of the intermediate states $\Gamma=0.3\ \rm eV$ was used to obtain the Chebyshev expansion degree as $K_G(\lambda)= \lambda\cdot(472+91\log\lambda)$ (see \cref{app:circuits} for more details of how this factor was obtained). The success probability $P_R$ was estimated classically for increasingly large active spaces using simulations in the previous section: it was observed to attain a stable value around $\sqrt{P_R}\approx 0.06$, which was used for calculating the associated $K_A=13$ rounds of amplitude amplification as shown in Eq.\eqref{eq:ka}. From this analysis, we find the cost of preparing the RIXS state requires about $40$ times more calls to the walk operator when compared to the QPE portion of the algorithm.

As seen in \cref{tab:re_rixs}, the dimension of the FCI matrix grows exponentially as we increase the number of orbitals and electrons, leading to the associated simulation of RIXS spectra quickly becoming classically intractable. By contrast, the non-Clifford gate counts for the quantum simulation are extremely attractive, having an increase of less than an order of magnitude as the number of orbitals was tripled from $10$ to $30$, with an associated approximate numerical scaling of $\mathcal{O}(N_a^{2.03})$.

While our resource estimates focus on a single polarization pair $(\bm\varepsilon_I,\bm\varepsilon_S)$, experimental spectroscopy of molecular clusters typically yields orientation-averaged quantities due to inhomogeneous molecular alignment \cite{mukamel}. Orientation invariant ``polarized'' and ``depolarized'' contributions are obtained from nine polarization combinations, $\{R_{ab}\}_{a,b\in\{x,y,z\}}$. Because each combination contributes only fractionally to the total spectrum, they can be computed with lower individual precision without compromising the final spectral accuracy. Consequently, by distributing QPE repetitions across these combinations according to their relative weights, the total cost for orientation-averaged RIXS spectra remains comparable to that of a single polarization pair.

\section{Conclusions}
\label{sec:conclusions}
Analysis of structural degradation issues in advanced battery cathodes like Li-excess requires applying a broad suite of spectroscopic methods, such as XAS and RIXS, both of which require accurate simulations for interpretation. Building on prior work targeting XAS spectra, in this manuscript we show how to use a quantum computer to simulate RIXS -- a complex, multidimensional, second-order spectroscopy -- in order to probe the low-lying excitations of cathode materials. Applying our performant qubitization-based spectroscopy workflow to a manganese-oxygen cluster identified in a recent RIXS study of Li-excess cathodes, we achieve RIXS simulation at only marginally larger cost than the (first-order) XAS process. The use of amplitude amplification to efficiently prepare the fixed-frequency incoming state allows us to alleviate one of the sampling bottlenecks, while a BLISS-THC-based qubitization approach ensures the circuit depth remains attractive.

Several avenues could be explored for further reduction of algorithmic complexity. First, a thorough characterization and fine-tuning of the qubitization hyperparameters to stay right at the accuracy threshold could feasibly yield lower per-step costs for specific systems, at the cost of only marginal extra classical computation. Second, demonstrating the applicability of various core-valence separation schemes, which are sometimes used in RIXS simulations, could allow to simplify the Hamiltonian and thus further reduce simulation costs, or even to introduce additional symmetry to be exploited by a BLISS-style approach. Finally, methods like interaction picture, or more recent techniques such as spectrum amplification via sum-of-squares factorizations could, if developed for this problem, also hold the prospect of improving the algorithm performance by decreasing the effective 1-norm of the Hamiltonian.

Overall, the fully specified quantum application of RIXS presented here is the first example of applying quantum algorithms to simulating a multidimensional, second-order spectroscopy in an industrially relevant context of battery research. Combined with algorithms for XAS, this study paves the way towards full readiness to deploying utility-scale fault tolerant quantum hardware for key R\&D challenges in the battery industry.

\acknowledgements
S.F, J.M.A. and A.D. acknowledge M.-L. Doublet and B. Senjean for fruitful discussions. The authors thank  Tarik El-Khateeb for producing the~\cref{fig:hero}. We acknowledge the Applied Quantum Computing Challenge program of the National Research Council of Canada for financial support (grant number AQC-103-2).

\bibliography{biblio}

\appendix
\onecolumngrid
\newpage

\section{Algorithmic details} \label{app:circuits}
\subsection{Green's function implementation}
\begin{figure}[]
   \centering
   \resizebox{1\textwidth}{!}{
   \begin{quantikz}
       \lstick{$\ket{0}$} && \gate[2]{\textrm{GQSP}(\hat{\mathcal{W}},K)} & \\
       \lstick{$\ket{\psi}$} & \qwbundle{2N_a} &&
   \end{quantikz}
   $=$ 
   \begin{quantikz}
	\lstick{$\ket{0}$} & \gate{R_z(\Phi)} & \gate{R_0} & \octrl{1} & \gate{R_1} & \ctrl{1} & \ \ldots \ & \octrl{1} & \gate{R_{2K-1}} & \ctrl{1} & \gate{R_{2K}} & \\
	\lstick{$\ket{\psi}$} & \qwbundle{2N_a} && \gate{\hat{\mathcal{W}}} && \gate{\hat{\mathcal{W}}^\dagger} & \ \ldots \ & \gate{\hat{\mathcal{W}}} && \gate{\hat{\mathcal{W}^\dagger}} &&  
\end{quantikz}}
   \caption{Green's function block-encoded implementation using generalized quantum signal processing (GQSP) on the qubitized walk operator $\hat{\mathcal{W}}$, with the top qubit being on $\ket{0}$ marking a successful application. Here we have used $R_k=R_z(\theta_k)R_y(\phi_k)$ to represent the single qubit rotations with associated GQSP angles $\{\theta_k,\phi_k\}$. For simplicity we do not explicitly show ancilla qubits used by walk operator. By convention we consider applications of $\hat{\mathcal{W}}$ and its hermitian conjugate to be alternating.}
   \label{fig:gqsp}
\end{figure}
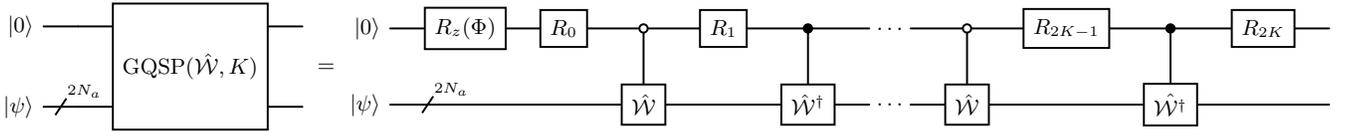

Here we discuss the details for block-encoding the Green's function in Eq.\eqref{eq:cheby_green} through the use of generalized quantum signal processing (GQSP) to implement its Chebyshev polynomial approximation
\begin{align}
    \Gamma\hat G(\omega_I,\Gamma) &\approx \sum_{k=0}^{K_G} c_k(\omega_I,\Gamma,\lambda)\cdot T_k\left(\frac{\hat H}{\lambda}\right) \label{eq:cheby_green_ap} \\
    &= \sum_{k=-K_G}^{K_G} \tilde c_k \hat{\mathcal{W}}^k,
\end{align}
where we have defined $\tilde c_0 = c_0$ and $\tilde c_k = \tilde c_{-k} = c_k/2$ for $k=1,\cdots,K_G$. The circuit in \cref{fig:gqsp} constitutes the application of GQSP, where the set of angles $\{\theta_k,\phi_k,\Phi\}$ define the associated $\rm SU(2)$ single-qubit rotations and can be found through a variety of efficient approaches \cite{gqsp,gqsp_angles,qsp_angles} once the Chebyshev expansion coefficients $\{c_k\}$ have been determined. \\

\begin{figure}
    \centering
    \includegraphics[width=0.65\linewidth]{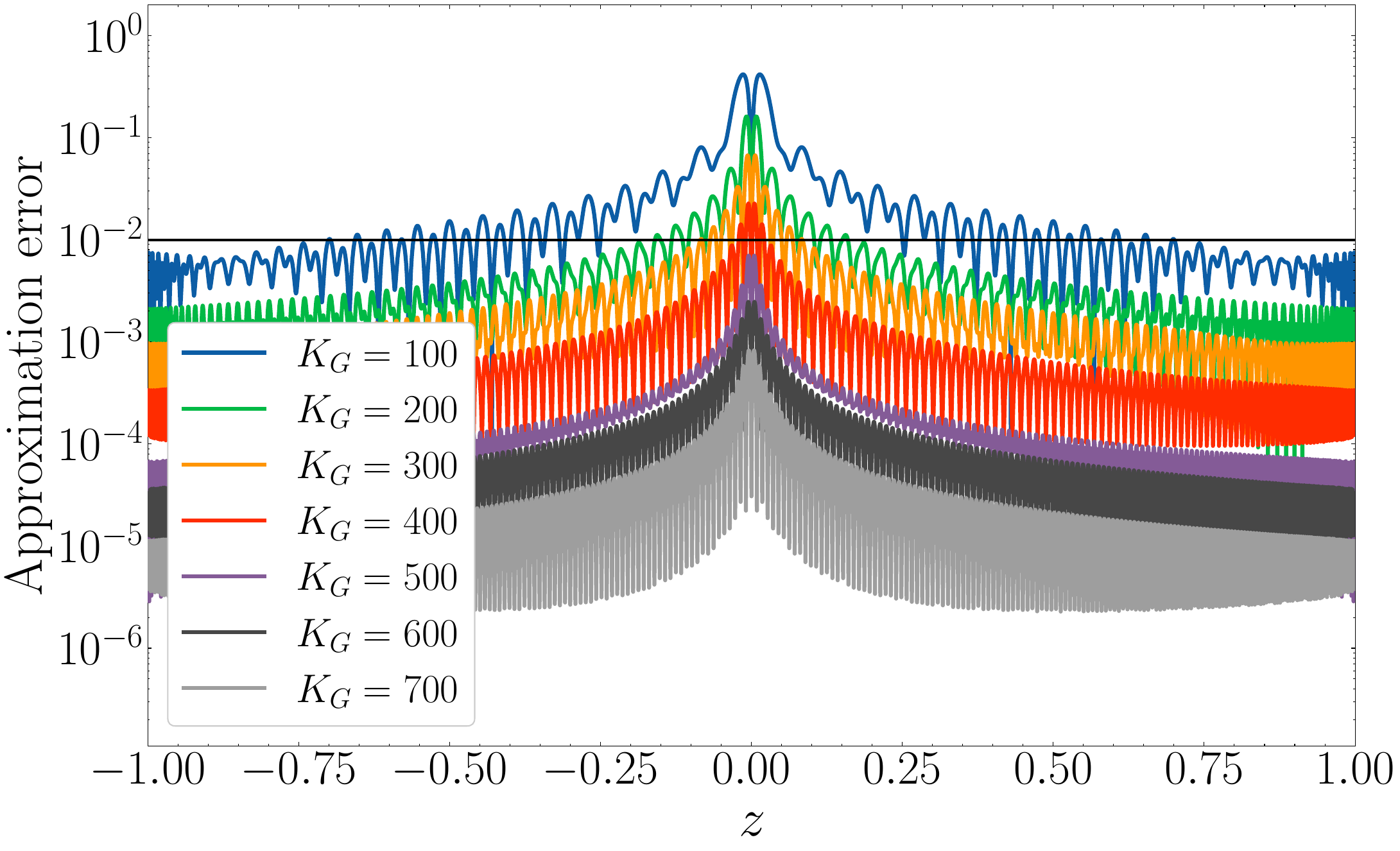}
    \caption{Absolute errors in Chebyshev expansion for different polynomial degrees $K_G$ considering $\lambda=1$ with the broadening $\Gamma=0.3\rm \ eV$.}
    \label{fig:cheby_error}
\end{figure}

We now discuss our choice of the Chebyshev expansion degree $K_G$, and provide a proof of its $\tilde{\mathcal{O}}(\lambda)$ scaling with respect to the 1-norm $\lambda$. We start by using the well-known result that the error in the $K_G$th Chebyshev approximation of some function $f(x)$ that is normalized (i.e. $|f(x)|\leq 1$) can be bounded with a geometric series as
\begin{equation}
    \epsilon(K_G) \leq \frac{2}{(\rho-1)\cdot\rho^{K_G}},
\end{equation}
with $\rho$ being the sum of the semi-axes of the Bernstein ellipse, namely the largest ellipse with foci at $\pm 1$ containing no singularities of $f(z)$ \cite{cheby_approx, cheby_approx_2, tang2023csguidequantumsingular}.Thus, the polynomial degree required for some target error $\epsilon$ is given by
\begin{equation}\label{eq:degree}
    K_G = \frac{1}{\log \rho} \cdot \left(\log \left(\frac{2}{\epsilon}\right)-\log(\rho-1)\right).
\end{equation}
We now consider the worst-case scenario for the Green's function where the ellipse will have the smallest $\rho$, which corresponds to the $\omega_I+E_0=0$ case with the associated target function 
\begin{equation}
    f(z;\lambda) = \frac{\Gamma}{i\Gamma -\lambda z},
\end{equation}
noting that the $\lambda$ factor here appears as to cancel the $\hat H/\lambda$ scaling from the walk operator. The associated complex pole is $z_p = i\Gamma/\lambda\equiv i\delta$, from which we can obtain the parameter $\rho$ for the Bernstein ellipse that touches this singularity
\begin{align}
    \rho &= |z_p |+ \sqrt{|z_p|^2+1} \\
    &= \delta +\sqrt{\delta^2+1}  \\
    &= 1 + \delta + \mathcal{O}\left(\delta^2\right),
\end{align}
having that the first line follows immediately from the definition of $\rho$ as the sum of semi-axes of the ellipse with foci at $\pm 1$ touching a purely imaginary pole $z_p$. Using the expansion $\log(1+x) = x+\mathcal{O}(x^2)$ alongside the fact that $\delta \ll 1$, we get $\log\rho\approx \delta\approx\rho-1$. Then, ~\cref{eq:degree} writes as follows
\begin{align}
    K_G &\approx \frac{1}{\delta} \cdot \left(\log \left(\frac{2}{\epsilon}\right) - \log\delta \right) \\
    &= \frac{\lambda}{\Gamma} \cdot \left(\log\left(\frac{2}{\epsilon}\right)+\log\left(\frac{\lambda}{\Gamma}\right)\right) \in \tilde{\mathcal{O}}(\lambda), \label{eq:cheby_eps_deg}
\end{align}
showing how the expansion degree $K_G$ grows linearly, up to logarithmic factors, with respect to the 1-norm of the Hamiltonian. In practice, we note that the error that is estimated from this analysis ends up being significantly smaller than the error we get from directly calculating the difference between $f(z)$ and its polynomial approximation. As such, we use the derived complexity here while doing a numerical fit for the $\epsilon$-dependent prefactor by considering the case of $\lambda=1$. From~\cref{fig:cheby_error} we determined that a polynomial degree of $472$ was enough to achieve errors smaller than $1\%$ across the full range. This is in contrast to the degree of $890$ that is obtained when using $\epsilon=10^{-2}$ in Eq.\eqref{eq:cheby_eps_deg}. Using $1/\Gamma\approx 91$ (in atomic units) for the considered broadening of $\Gamma=0.3\rm \ eV$, we obtain that the required polynomial expansion degree as a function of the 1-norm follows the equation
\begin{equation}
    K_G(\lambda) = \ceil{\lambda\cdot(472+91\log\lambda)}.
\end{equation}

\subsection{Dipole block-encoding}
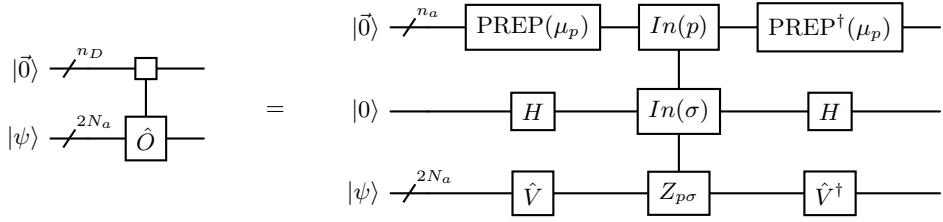
\begin{figure}
    \centering
    \begin{quantikz}
        \lstick{$\ket{\vec 0}$} & \qwbundle{n_D} & \gate{} \vqw{1} & \\
        \lstick{$\ket{\psi}$} & \qwbundle{2N_a} & \gate{\hat O} &
    \end{quantikz}
   \hspace{0.5cm} $=$ \hspace{0.5cm}
    \begin{quantikz}
            \lstick{$\ket{\vec 0}$} & \qwbundle{n_a} & \gate{\textup{PREP}(\mu_p)} & \gate{In(p)} \vqw{2} & \gate{\textup{PREP}^\dagger(\mu_p)} & \\
            \lstick{$\ket{0}$} && \gate{H} & \gate{In(\sigma)} & \gate{H} & \\
            \lstick{$\ket{\psi}$} & \qwbundle{2N_a} & \gate{\hat V} & \gate{Z_{p\sigma}} & \gate{\hat V^\dagger} &
    \end{quantikz}
    \caption{Block-encoding of dipole operator $\hat O$ in Eq.\eqref{eq:dip_trunc} without identity term. The $\textup{PREP}(\mu_p)$ circuit here uses coherent alias sampling \cite{qrom} to load the associated coefficients, while $In(p)$ corresponds to a unary iteration \cite{qrom} and the orbital rotation $\hat V$ is implemented using Givens rotations \cite{df}. Note that dirty ancillas generated by the coherent alias sampling routine should be used during the implementation of its hermitian conjugate. All $n_D$ ancilla qubits are returned in the zero state upon a successful application of the block-encoding.}
    \label{fig:dip_be}
\end{figure}

For block-encoding the dipole operator, we here present a general strategy for block-encoding one-electron observables with an optimal 1-norm, as discussed in Ref. \cite{loaiza_lcu}, which we apply to the dipole operator in Eq.\eqref{eq:dip_cvs}. Start by considering a general one-electron (spin-conserving) hermitian operator, which can be diagonalized via an orbital rotation $\hat V$ as
\begin{align}
    \hat O &= \sum_{pq,\sigma} o_{pq} \hat c^\dagger_{p\sigma} \hat c_{q\sigma} \label{eq:dip_trunc} \\
    &= \sum_{p,\sigma} \mu_{p} \sum_{qs} V_{pq} V_{ps} \hat c_{q\sigma}^\dagger \hat c_{s\sigma}  \\
    &= \sum_{p,\sigma} \frac{\mu_p}{2} \hat V^\dagger\hat z_{p\sigma} \hat V + \hat 1\sum_{p} \mu_p,
\end{align}
where we have used the fact that the number operator $\hat c_{p\sigma}^\dagger \hat c_{p\sigma}$ is mapped to $(\hat 1 + \hat z_{p\sigma})/2$ for $\hat z_{p\sigma}$ under standard fermion-to-qubit mappings such as Jordan-Wigner or Bravyi-Kitaev \cite{jw_1,jw_2,bk_1,bk2}, and $\mu_\rho$ correspond to the eigenvalues obtained when diagonalizing the $[o_{pq}]$ matrix. Here $\hat z_{p\sigma}$ is the Pauli Z operator acting on the qubit encoding spin-orbital $p\sigma$. We only consider the implementation of the non-identity component in the second line of this last expression, given how the identity component acts trivially while increasing the overall implementation cost. The associated 1-norm is given by
\begin{equation}
    \lambda_O = \sum_p |\mu_p|.
\end{equation}
This operator can be block-encoded by using a standard approach for linear combinations of unitaries via a PREPARE and SELECT circuits. The PREPARE circuit loads the $\mu_p$ coefficients via coherent alias sampling \cite{qrom}, while the SELECT circuit rotates to the $\hat V$ orbital frame while implementing the $\hat z_{p\sigma}$ Pauli Z operators by using a unary iteration \cite{qrom} running over the different orbital indices $p$. An orbital rotation acting on $N_a$ orbitals can be generally implemented by decomposing it into $N_a(N_a-1)/2$ Givens rotations \cite{df}. The associated circuit for block-encoding $\hat O$ is shown in \cref{fig:dip_be}. Overall, this routine will require the following ancilla qubits besides the $2N_a$ system qubits, where we use the shorthand notation $n_a \equiv \ceil{\log_2 N_a}$ (noting that $\ceil{\log_2(2N_a)} = n_a+1$):
\begin{itemize}
    \item $n_a$ qubits for the unary iteration routine over $2N_a$ different values of $p$. 
    \item $n_a$ qubits for storing the $N_a$ different values of $\mu_p$.
    \item $1$ qubit for encoding the spin index $\sigma$.
    \item $2\aleph_{\mu} + n_a + 1$ qubits for the coherent alias sampling routine \cite{qrom}, where $\aleph_\mu$ is the number of qubits used for the precision of the associated $\mu_p$ coefficients that are loaded and $n_a$ qubits are used for the $\rm alt$ register. Note that we do not consider the additional qubits required by the QROM in the coherent alias sampling as these can be shared with those used by the unary iteration. 
\end{itemize}
Overall, this routine uses
\begin{align}
    n_D &= 3n_a + 2\aleph_\mu+2
\end{align}
ancilla qubits.

\section{Active space selection}
\label{app:avas}
To select the active orbitals for the $\text{MnO}_7\text{H}_6$ cluster, we applied AVAS with different thresholds for occupied ($\mathbf{th}_\text{o}$) and virtual ($\mathbf{th}_\text{v}$) portions of the active space. This was accomplished by fixing a combination of reference orbitals ($\mathtt{ao\_labels}$ parameter in the AVAS subroutine implemented in $\mathtt{PySCF}$) and the  ``minimal'' basis, diagonalizing $P_A$ on occupied and virtual ROHF subspaces, and applying eigenvalue thresholds listed in \cref{tab:avas_params}.

\begin{table}[h]
\centering
\begin{tabular}{c c c c c}
\hline
Number of electrons $(N_e)$ &~~~~ Number of orbitals ($N_a$) &~~~~$\mathbf{th}_\text{o}$ &~~~~$\mathbf{th}_\text{v}$  \\
\hline
$15$ &~~~~ $16$ &~~~~ $7.0\times10^{-1}$ &~~~~ $1.52 \times 10^{-1}$ \\
$19$ &~~~~ $18$ &~~~~ $5.0\times10^{-1}$ &~~~~ $4.00\times10^{-1}$ \\
$19$ &~~~~ $20$ &~~~~ $5.0\times10^{-1}$ &~~~~ $1.60\times10^{-1}$ \\
$21$ &~~~~ $22$ &~~~~ $4.0\times10^{-1}$ &~~~~ $1.60\times10^{-1}$ \\
$21$ &~~~~ $24$ &~~~~ $4.0\times10^{-1}$ &~~~~ $1.00\times10^{-1}$ \\
$21$ &~~~~ $26$ &~~~~ $4.0\times10^{-1}$ &~~~~ $6.00\times10^{-5}$ \\
$23$ &~~~~ $28$ &~~~~ $2.0\times10^{-1}$ &~~~~ $5.00\times10^{-5}$ \\
$27$ &~~~~ $30$ &~~~~ $1.0\times10^{-1}$ &~~~~ $5.00\times10^{-5}$ \\
\hline
\end{tabular}
\caption{AVAS threshold parameters used to build the active spaces in \cref{tab:re_rixs}.  Reference orbitals were represented in $6-31\text{G}$ basis to allow for the $4p$-like set on $\text{Mn}$. For $N_a <19$ $\mathtt{ao\_labels} = [\text{O}~1s, \text{O}~2p, \text{Mn}~3d]$, and for $N_a \geq 19$ $\mathtt{ao\_labels} = [\text{O}~1s, \text{O}~2p, \text{Mn}~3d, \text{Mn}~4p]$, where label $\text{O}$ refers to the peroxide group oxygen in \cref{fig:cluster}. }\label{tab:avas_params}
\end{table}

\end{document}